%% file: main.tex
\documentclass[twocolumn]{aastex63}

\usepackage{booktabs}
\usepackage[flushleft]{threeparttable}
\usepackage[utf8]{inputenc}
\usepackage{chngpage}
\usepackage{float}

\urldef{\footurl}\url{https://skaafrica.atlassian.net/wiki/spaces/ESDKB/pages/338723406/}

\usepackage{graphicx}	
\usepackage{amsmath}	
\usepackage{amssymb}	

\shorttitle{Radio Evolution of the TDE ASASSN-19bt}
\shortauthors{Christy et al.}
\graphicspath{{./}{figures/}}

\begin{document}

\title[ASASSN-19bt]{The Peculiar Radio Evolution of the Tidal Disruption Event ASASSN-19bt}

\input{authors}

\begin{abstract}
\indent We present detailed radio observations of the tidal disruption event (TDE) ASASSN-19bt/AT2019ahk, obtained with the Australia Telescope Compact Array (ATCA), the Atacama Large Millimeter/submillimeter Array (ALMA), and the MeerKAT radio telescopes, spanning 40 to 1464 days after the onset of the optical flare. We find that ASASSN-19bt displays unusual radio evolution compared to other TDEs, as the peak brightness of its radio emission increases rapidly until 457 days post-optical discovery and then plateaus. Using a generalized approach to standard equipartition techniques, we estimate the energy and corresponding physical parameters for two possible emission geometries: a non-relativistic spherical outflow and a relativistic outflow observed from an arbitrary viewing angle. We find that the non-relativistic solution implies a continuous energy rise in the outflow from $E\sim10^{46}$ erg to $E\sim10^{49}$ erg with $\beta \approx 0.05$, while the off-axis relativistic jet solution instead suggests $E\approx10^{52}$ erg with $\Gamma\sim10$ erg at late times in the maximally off-axis case. We find that neither model provides a holistic explanation for the origin and evolution of the radio emission, emphasizing the need for more complex models. ASASSN-19bt joins the population of TDEs that display unusual radio emission at late times. Conducting long-term radio observations of these TDEs, especially during the later phases, will be crucial for understanding how these types of radio emission in TDEs are produced.

\end{abstract}



\section{Introduction}

Tidal disruption events (TDEs) arise as a natural consequence of having supermassive black holes (SMBHs) populate the centers of galaxies. These events occur when a star's orbit passes within the tidal radius of a massive black hole; for solar-type stars, this occurs outside of the event horizon for black holes with masses $M_{\rm BH} \lesssim 10^8 M_{\odot}$ \citep{hills_1975}. When a star passes close enough to be disrupted in a single flyby, roughly half of the debris escapes on hyperbolic orbits, while the remaining material circularizes to form an accretion disk around the black hole \citep{Rees1998}. TDEs are important to study because they act as cosmic laboratories for accretion physics, outflow mechanisms, and jet formation. Radio observations in particular are used to characterize outflows in TDEs and other extragalactic transient phenomena \citep{alexander2020}. Radio observations of transients provide key diagnostics such as calorimetry, outflow velocity, magnetic field strength, and the density of the immediate environments surrounding the transient \citep[e.g.,][]{1998chevalier,Metzger_2012,Margalit_2021,matsumoto}.

In recent years, many TDEs have undergone follow-up radio observations, revealing a wide range of outflow properties (see \citealt{alexander2020} for a review). A few TDEs have been observed to launch luminous, on-axis relativistic jets \citep[e.g.,][]{Zauderer_2011, cenko_2012, Brown_2017, Andreoni_2022, AT2022cmc_pasham_2023, AT2022cmc_matsumoto_metzger2023, Yao_2024} showing long-lasting luminous radio emission after the initial disruption of the star. More commonly, TDEs may exhibit fainter radio emission consistent with a non-relativistic outflow from accretion-driven winds, the unbound debris expanding into the surrounding medium, and/or collision-induced outflows from the fallback stream \citep[e.g.,][]{Strubbe, Tchekhovskoy, alexander_2016, krolik_2016, bonnerot_2020, at2019dsg}. However, the exact origin of the less luminous radio emission is not yet clear; any/multiple of these possibilities may be correct. It is also common for follow-up observations of TDEs to reveal no radio emission at all on timescales of days to years post-disruption, which implies the absence of an energetic outflow with emission pointing along our line of sight \citep{alexander2020, van_Velzen_2023}. 

Recently, as many as 40\% of TDEs have been reported to show delayed ($\gtrsim 1$ yr) radio emission relative to their optical discovery, suggesting that previous radio monitoring campaigns were insufficiently sensitive or ended too early \citep{yvette_2023}. An early example of such a case is the TDE ASASSN-15oi, which displayed no prompt radio emission, then a detection of a flare six months later, followed by a second and brighter flare years later \citep{horesh_15oi}. Another notable example is the TDE AT2018hyz which showed rapidly rising radio emission after two years of non-detections \citep{AT2018hyz, Sfaradi_2024}. In a few instances, delayed low luminosity radio emission in TDE candidates has been successfully modeled as off-axis jets \citep{perlman_2017, ARP_2018}, but the overall prevalence of jets in TDEs remains an open question hampered by inconsistent radio follow up. Radio emission from a relativistic jet viewed off-axis may not be visible until months or years post-disruption when the jet decelerates, and in such cases careful multi-frequency modeling spanning several years may be required to discriminate between an off-axis jet and a delayed non-relativistic outflow \citep{matsumoto}. Thus, dedicated long-term radio monitoring campaigns are necessary to fully characterize the origin(s) of radio emission in individual TDEs.

Adding to the existing set of TDE observations, we present $\sim$ 4 years of radio observations of the TDE ASASSN-19bt/AT2019ahk. ASASSN-19bt ($z = 0.0262$) was discovered by the All-Sky Automated Survey for SuperNovae \citep[ASAS-SN;][]{Shappee, Kochanek_2017} on 2019 January 29 in the galaxy 2MASX J07001137-6602251 \citep{holoien_2019}. This event marked the first TDE discovered in the Transiting Exoplanet Survey Satellite \citep[TESS;][]{Ricker_2015} Continuous Viewing Zone, resulting in a high-cadence optical light curve. Upon this source's classification as a TDE, we quickly began our radio monitoring, with our first observation occurring near the time of peak optical emission. In their discovery paper, \citet{holoien_2019} found that the optical/UV emission of ASASSN-19bt indicated a luminosity, temperature, radius, and spectroscopic evolution similar to those of previously-studied optical TDEs. However, its X-ray properties are unusual: its X-ray luminosity is among the lowest observed for any optically-selected TDE, and \citet{holoien_2019} report an X-ray photon index of $\Gamma = 1.47^{+0.3}_{-0.3}$ at the time of peak optical light and $\Gamma = 2.34^{+0.8}_{-0.6}$ after optical peak, both harder than those of many optically-selected TDEs \citep{guolo_23}.  Radio observations may illuminate the nature of this emission by constraining the disruption geometry and the characteristics of any resulting outflows.


In this work, we present the radio evolution of ASASSN-19bt and consider two possible models: an off-axis relativistic jet or a spherical non-relativistic outflow. We find that the emission cannot be fully described by either of these simple limiting cases. 
In \S \ref{sec:obs}, we outline our observations and data reduction methods. In \S \ref{sec:modeling},  we describe our modeling framework for the radio emission. In \S \ref{sec:analysis}, we infer the physical properties of the outflow and the circumnuclear density for both models. 
In \S \ref{sec:disc}, we examine the characteristics of the outflow and discuss ASASSN-19bt in context of other TDEs. In \S \ref{sec:conc} we summarize our findings and present our conclusions.

\section{Observations} \label{sec:obs}
We obtained radio, mm, and X-ray observations of ASASSN-19bt as described below. In addition, we also use archival X-ray and radio data and data first published in \citet{holoien_2019}. We define $\delta t$ as the time between our observation and 2019 January 21.6, the date where the TESS light curve indicates that the transient began to brighten \citep{holoien_2019}. We assume a distance of 115.2 Mpc for ASASSN-19bt.

\begin{table*}
\centering
\caption{Radio observations of ASASSN-19bt. We report the uncertainties as 1$\sigma$ statistical error with an additional 5\% systematic error term to account for uncertainties in the absolute flux density scale calibration. Non-detections are reported as 3$\sigma$ upper limits. All values of $\delta t$ are relative to 2019 January 21.6 \citep{holoien_2019}.}
\label{tab:radio_data}
    \begin{tabular}{llcccc}
    \hline\hline
    Date  &  Telescope     &  Array   &  $\delta t$  &  $\nu$       &  $F_\nu$ ($\mu$Jy)  \\
    (UTC)  &                &      Configuration        &  (days)      &  (GHz)       &  $\pm$ statistical error \\ 
      &                &              &              &              &  $\pm$ systematic error  \\
    \hline
    2015-12-07   &   ATCA       & 1.5A       & -1142      & 2.1        &   464 $\pm $ 70 $\pm$ 23   \\ 
    2016-08-19   &   ATCA       & 6C         & -886       & 2.1        &   400 $\pm $ 86 $\pm $ 20   \\ 
    \hline
    2019-03-03   &   ALMA       & C43-1      & 40         & 97.5       &   143 $\pm $ 21 $\pm $ 7   \\ 
    2019-03-12   &   ATCA       & 6A         & 49         & 16.7       &   220 $\pm $ 13 $\pm $ 11  \\ 
    2019-03-12   &   ATCA       & 6A         & 49         & 21.2       &   206 $\pm $ 23 $\pm $ 10  \\ 
    2019-03-24   &   ALMA       & C43-2      & 61         & 97.5       &   89 $\pm $ 16 $\pm $ 4  \\ 
    \hline
    2019-04-11   &   ATCA       & 750C       & 79         & 17.0       &   235 $\pm $ 25 $\pm $ 11  \\ 
    2019-04-11   &   ATCA       & 750C       & 79         & 19.0       &   216 $\pm $ 28 $\pm $ 10  \\ 
    2019-04-13   &   ALMA       & C43-3      & 81         & 97.5       &   60 $\pm $ 14 $\pm $ 3  \\ 
    2019-04-19   &   ATCA       & 750C       & 87         & 5.5        &   418 $\pm $ 32 $\pm $ 20  \\ 
    2019-04-19   &   ATCA       & 750C       & 87         & 9.0        &   330 $\pm $ 25 $\pm $ 16  \\ 
    2019-05-15   &   ATCA       & 1.5B       & 113        & 2.1        &   440 $\pm $ 60 $\pm $ 22  \\ 
    \hline
    2019-06-03   &   ATCA       & 6A         & 132        & 2.1        &   789 $\pm $ 91 $\pm $ 39  \\ 
    2019-06-03   &   ATCA       & 6A         & 132        & 5.5        &   630 $\pm $ 43 $\pm $ 31  \\ 
    2019-06-03   &   ATCA       & 6A         & 132        & 9.0        &   493 $\pm $ 40 $\pm $ 24  \\ 
    2019-06-03   &   ATCA       & 6A         & 132        & 17.0       &   276 $\pm $ 32 $\pm $ 13  \\ 
    2019-06-03   &   ATCA       & 6A         & 132        & 19.0       &   266 $\pm $ 37 $\pm $ 13  \\ 
    2019-06-20   &   ALMA       & C43-9/10   & 149        & 97.5       &   $<$31     \\ 
    \hline
    2019-07-27   &   ATCA       & 750C       & 186        & 2.1        &   1153 $\pm $ 125 $\pm $ 57  \\ 
    2019-07-27   &   ATCA       & 750C       & 186        & 5.5        &   759 $\pm $ 35 $\pm $ 37  \\ 
    2019-07-27   &   ATCA       & 750C       & 186        & 9.0        &   497 $\pm $ 25 $\pm $ 24  \\ 
    2019-07-27   &   ATCA       & 750C       & 186        & 17.0       &   272 $\pm $ 34 $\pm $ 13  \\ 
    2019-07-27   &   ATCA       & 750C       & 186        & 19.0       &   194 $\pm $ 25 $\pm $ 9  \\ 
    \hline
    2020-04-23   &   ATCA       & 6A         & 457        & 2.1        &   3589 $\pm $ 127 $\pm $ 179  \\ 
    2020-04-23   &   ATCA       & 6A         & 457        & 5.0        &   5044 $\pm $ 91 $\pm $ 252  \\ 
    2020-04-23   &   ATCA       & 6A         & 457        & 6.0        &   4833 $\pm $ 40 $\pm $ 241  \\ 
    2020-04-23   &   ATCA       & 6A         & 457        & 8.5        &   3907 $\pm $ 38 $\pm $ 195  \\ 
    2020-04-23   &   ATCA       & 6A         & 457        & 9.5        &   3453 $\pm $ 40 $\pm $ 172  \\ 
    2020-04-23   &   ATCA       & 6A         & 457        & 17.0       &   1806 $\pm $ 115 $\pm $ 90  \\ 
    2020-04-23   &   ATCA       & 6A         & 457        & 19.0       &   1494 $\pm $ 154 $\pm $ 74  \\ 
    \hline
    2021-03-25   &   ATCA       & 6D         & 793        & 5.0        &   3568 $\pm $ 154 $\pm $ 178  \\ 
    2021-03-25   &   ATCA       & 6D         & 793        & 6.0        &   3078 $\pm $ 74 $\pm $ 153  \\ 
    2021-03-25   &   ATCA       & 6D         & 793        & 8.5        &   1709 $\pm $ 83 $\pm $ 85  \\ 
    2021-03-25   &   ATCA       & 6D         & 793        & 9.5        &   1265 $\pm $ 94 $\pm $ 63  \\ 
    2021-04-10   &   ATCA       & 6D         & 809        & 2.1        &   5000 $\pm $ 250 $\pm $ 250  \\ 
    2021-04-19   &   MeerKAT    & -          & 818        & 1.3        &   4400 $\pm $ 140 $\pm $ 220  \\ 
    \hline
    2022-09-30   &   ATCA       & 6D         & 1347       & 1.6        &   5503 $\pm $ 140 $\pm $ 275  \\ 
    2022-09-30   &   ATCA       & 6D         & 1347       & 2.6        &   4015 $\pm $ 89 $\pm $ 200  \\ 
    2022-09-30   &   ATCA       & 6D         & 1347       & 5.0        &   2418 $\pm $ 54 $\pm $ 120  \\ 
    2022-09-30   &   ATCA       & 6D         & 1347       & 6.0        &   2087 $\pm $ 35 $\pm $ 104  \\ 
    2022-09-30   &   ATCA       & 6D         & 1347       & 8.5        &   1414 $\pm $ 37 $\pm $ 70  \\ 
    2022-09-30   &   ATCA       & 6D         & 1347       & 9.5        &   1299 $\pm $ 40 $\pm $ 64  \\ 
    2023-01-25   &   MeerKAT    & -          & 1464       & 0.8        &   6764 $\pm $ 54 $\pm $ 338  \\ 
    2023-01-25   &   MeerKAT    & -          & 1464       & 1.3        &   6145 $\pm $ 39 $\pm $ 307  \\ 
        \hline \hline
\end{tabular}
\end{table*}

\subsection{Radio/Millimeter}


We collected radio observations of ASASSN-19bt using the Australia Telescope Compact Array (ATCA) in the 6A, 6D, 1.5B, 750C configurations with a frequency range 2-19 GHz under programs C3325 (PI: K. Alexander) and CX432 (Director's Time, PI: Alexander). Additionally, we included data from 2019-03-12 under program C3148 (PI: S. van Velzen). The results of these observations can be found in Table \ref{tab:radio_data}. To process the data, we applied standard data reduction techniques using the package \textit{Miriad} \citep{Miriad}. For calibration, we employed B1934-638 as the primary flux calibrator for all observations and frequencies. J0700-6610 served as the phase calibrator for all frequencies. For the 2.1 GHz observations, we centered the field on J0700-6610 rather than slewing back and forth to the target. Due to its proximity to the TDE, we employed the ``peeling'' technique to improve the data calibration (see e.g., \citealt{peeling}). This is a procedure for applying direction dependent gains to a limited portion of an image to reduce the artifacts due to a nearby bright source.

The data were imaged using the \textit{Miriad} tasks \verb'mfclean,  restor'. In observations where the target was sufficiently bright, we split the data into 2 subbands of 1024 MHz for imaging. Flux densities and their associated uncertainties were determined by fitting a point source model (gaussian with the width of the point-spread-function) using the \verb'imfit' command within \textit{Miriad}. 

We also obtained four Band 3 observations using the Atacama Large Millimeter/submillimeter Array (ALMA) with a mean frequency of 97.5 GHz under program 2018.1.01766.T (PI: K. Alexander). The data were calibrated with the ALMA data pipeline version Pipeline-CASA54-P1-B using J0519-4546 as the flux calibrator and J0700-6610 as the phase calibrator. We detected ASASSN-19bt in the first three observations, while the 2019 June 20 observation yielded an upper limit. We imaged the data using the Common Astronomy Software Applications (CASA; version 6.2.1.7) software package \citep{CASA_team}, wherein we fit the flux densities and associated uncertainties by fitting an elliptical Gaussian fixed to the size of the synthesized beam using the CASA task \verb'imfit'. 

Finally, we obtained observations with MeerKAT at 0.82 GHz ($UHF-$ band) and 1.28 GHz ($L-$band) under program SCI-20210212-YC-01 (PI: Y. Cendes). For these observations we used J0408-6545 as the flux calibrator and J0906-6829 as the phase calibrator. We used the calibrated images obtained via the South African Radio Astronomy Observatory Science Data Processor (SDP) \footnote{\footurl}. For these observations, we fit the flux densities and associated uncertainties by fitting an elliptical Gaussian fixed to the size of the synthesized beam using the CASA task \verb'imfit'. 

We note that the uncertainties derived from our fitting procedure are statistical errors only. When modelling we include an additional uncertainty of 5\% of the source flux density to account for the known absolute flux density scale calibration accuracy of ATCA, ALMA, and MeerKAT. We include both sets of errors in Table \ref{tab:radio_data}.

\begin{figure}
    \centering
    \includegraphics[width = 0.5\textwidth]{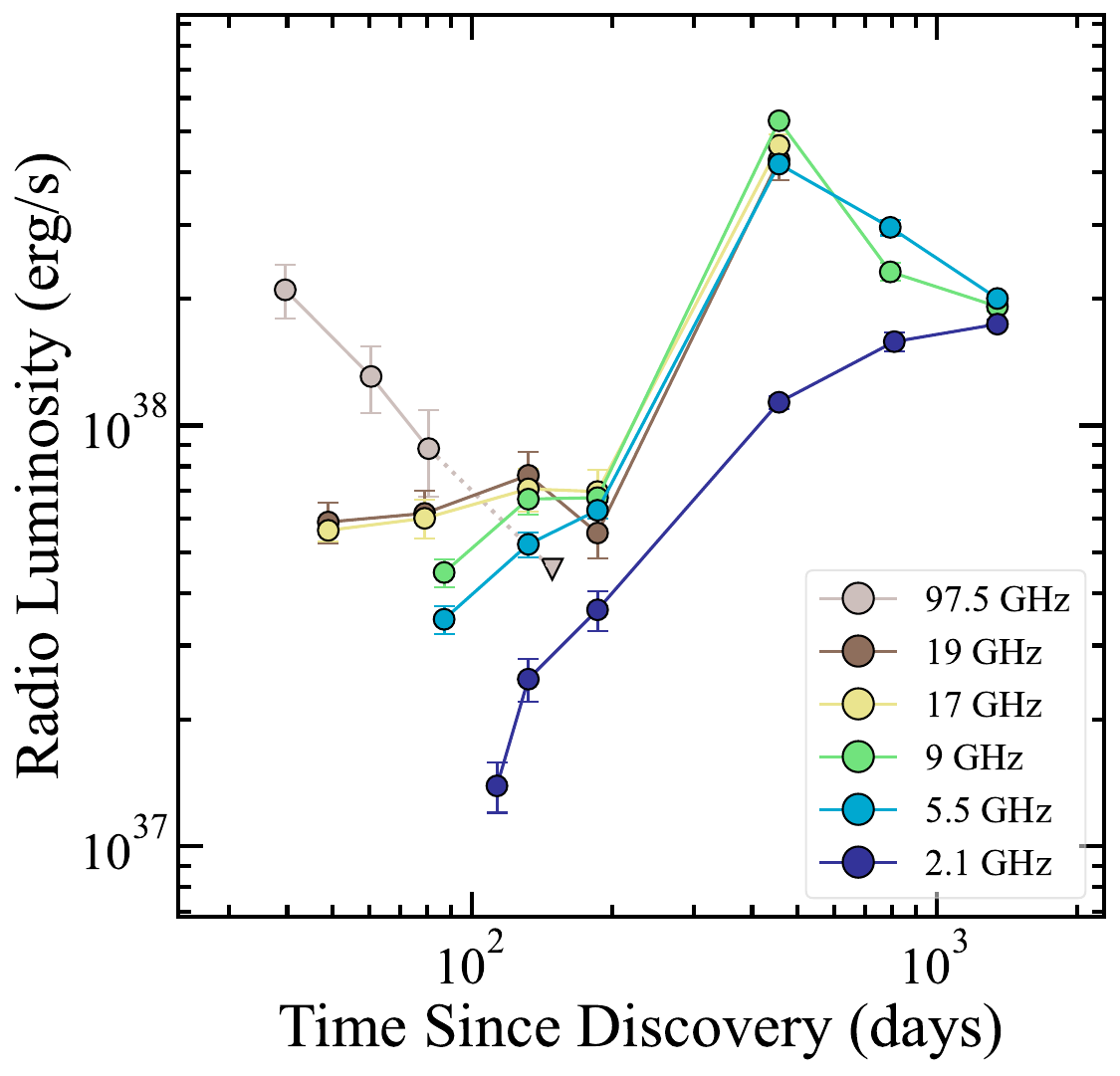}
    \caption{Radio and millimeter light curves of ASASSN-19bt, including an upper limit (3$\sigma$) from the last ALMA observation at 97.5 GHz. Our earliest radio detections occurred $\sim$2 days before peak optical light, constraining the outflow launch time to before this date. The source brightens significantly in all observed ATCA frequencies (2.1-19 GHz) $\sim1$ year after the optical discovery.}
    \label{fig:radio_lightcurves}
\end{figure}

\begin{figure*}[ht!]
    \centering
    \includegraphics[width = \textwidth]{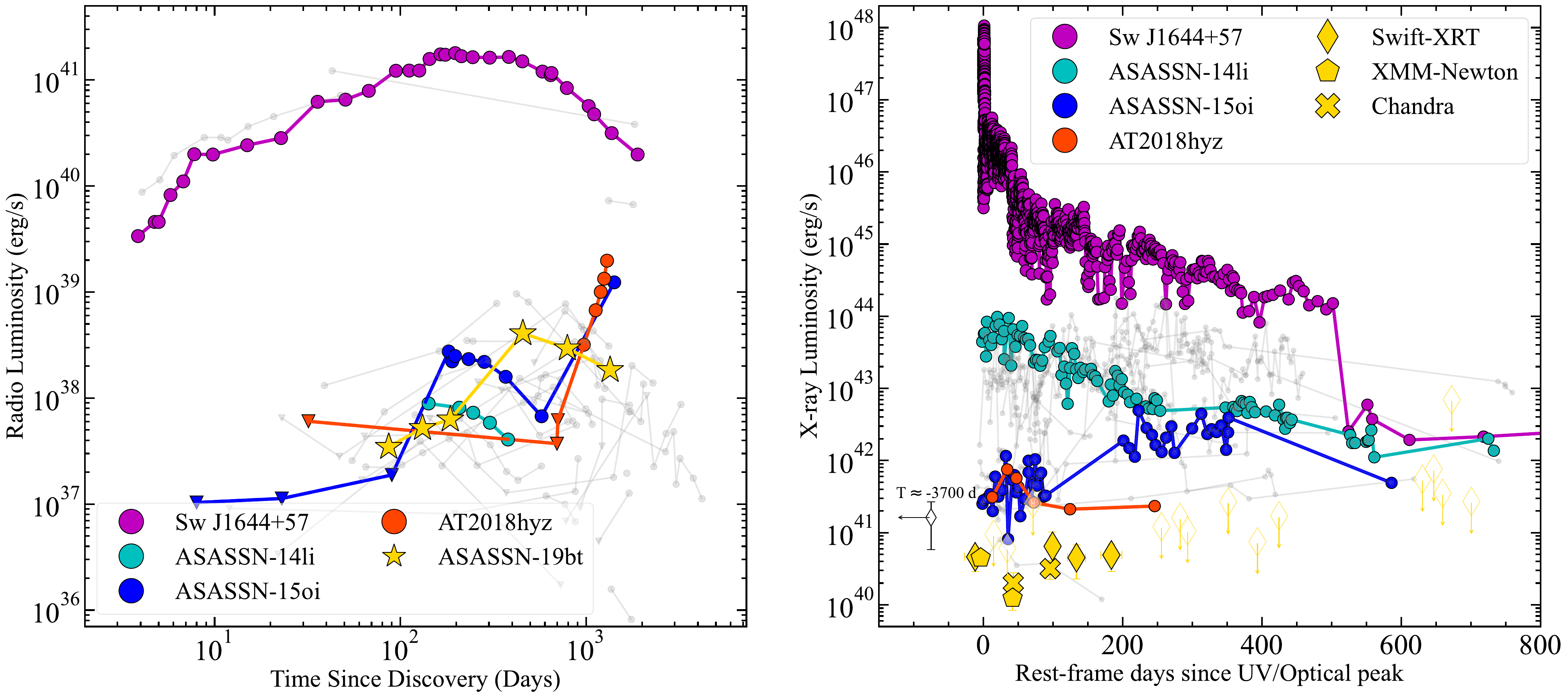}
    \caption{\textit{Left:} Radio luminosity of ASASSN-19bt at C band ($\approx5.5$ GHz) in comparison to a selection of previously studied TDEs with radio observations. Highlighted TDE data are from Sw J1644+57 \citep{Zauderer_2011,Eftekhari}, ASASSN-14li \citep{alexander_2016}, ASASSN-15oi \citep{horesh_15oi}, and AT2018hyz \citep{AT2018hyz}. Gray points are data from the literature \citep[and references therein]{alexander2020, yvette_2023} \textit{Right:} X-ray luminosity of ASASSN-19bt in comparison to the sample of X-ray light curves from \citet{guolo_23}. While ASASSN-19bt's radio luminosity is typical of the overall population, it has some of the faintest known X-ray emission of any TDE at both early and late times (yellow symbols). The marginal archival X-ray detection coincident with the nucleus of ASASSN-19bt's host galaxy $\sim3700$ d before the TDE is shown with an open diamond.
    }
    \label{fig:lit_lc}
\end{figure*}

\subsection{Archival Radio Observations}\label{sec:archival_radio}
The host galaxy of ASASSN-19bt was previously detected by ATCA serendipitously due to its proximity to J0700-6610. These observations were taken at 2.1 GHz in 2015 December and 2016 August under the programs C1473 (PI: S. Ryder) and C3101 (PI: A. Edge) respectively. We reduced and imaged these data using the same procedure outlined above for our ATCA observations. The resulting archival flux densities are shown in Table \ref{tab:radio_data}. A pre-disruption radio detection implies that some of the radio emission we observe is due to background emission from the host galaxy and not associated with the TDE. If the host galaxy emission component is constant in time, then the archival radio observations and the non-detection from ALMA in 2019 require that the host spectral energy distribution (SED) at radio frequencies follows a power law $F_\nu\propto\nu^{-0.7}$ or steeper. 
For our modeling, we follow the same procedure outlined in \citet{alexander_2016}, and assume that the host component of the radio emission follows a single power law with $F_\nu\approx(439\ \rm{\mu Jy}) (\nu/2.1 \rm GHz)^{-1}$, where 439 $\rm{\mu Jy}$ is the weighted average of the two archival detections at 2.1 GHz. 

We find that the archival detection is consistent with radio emission from ongoing star formation in the host galaxy. Extrapolating the quiescent component to 1.4 GHz, the observed radio luminosity requires the formation rate of stars more massive than 5 $M_\odot$ to be $\sim0.22$ $M_\odot \ \rm yr^{-1}$ \citep{Condon_2002}. This is consistent with \citet{holoien_2019}, who reported that an archival spectrum of 2MASX J07001137-6602251 was well fit by a star-formation rate (SFR) of $0.17^{+0.06}_{-0.01}$ $M_\odot \ \rm yr^{-1}$. We note that \citet{holoien_2019} did not include a possible AGN component when modelling the archival host galaxy spectrum, therefore the inferred SFR should only be considered an upper limit.

Alternately, the pre-disruption radio detection could imply the presence of a compact AGN. 
The non-detection from ALMA suggests that if prior to the TDE the host contained a compact AGN with the flat radio spectrum typically seen in such sources, then its radio emission at all wavelengths must have been quenched at the time of disruption, and 100\% of the emission observed during our monitoring campaign originated from the TDE. 
A steep-spectrum AGN component is allowed by our data, but in this case the host emission would likely arise from a relic AGN jet at sufficiently large physical separation from the SMBH to be unaffected by the TDE. We thus conclude it is reasonable to assume that the host galaxy emission component is constant in time.


After subtracting the host component, we find that the resulting transient component exhibits the shape of a self-absorbed synchrotron spectrum in all epochs (see Figure \ref{fig:1}).  The remainder of our analysis will focus on modelling and interpreting the transient component of the radio emission. For completeness, we also model the observed emission as a single component and present our results in Appendix \ref{appendix:sed}; we note that these results do not alter the basic conclusions of our analysis.
\begin{table*}[!ht]
    \centering
    \caption{X-ray luminosity (0.3-10.0 keV) of ASASSN-19bt. The uncertainties denote the $1\sigma$ confidence interval while the upper limits are $3\sigma$ limits. We outline our reduction methods for the Chandra and Swift-XRT data in sections \S \ref{subsec:ChandraDataAnalysis},\ref{subsec:SwiftDataAnalysis}. For discussion of the XMM-Newton data reduction, see \citet{holoien_2019}.}
    \label{tab:xray}
    \begin{tabular}{lllcc}
    \hline\hline
        Instrument & Observation IDs & 	 MJD & 	 $\delta t$ & $L_X/10^{40}$ erg/s \\ 
        \hline
        Swift-XRT  & 38456001	  & 	 54853.65 $^{+0.17}_{-0.03}$ & 	 -3652  &  16.26$^{+12.44}_{-8.37}$  \\
        Swift-XRT  & 38456002	  & 	 54874.51 $^{+0.13}_{-0.13}$ & 	 -3631  &  $<$30.49  \\
        Swift-XRT  & 41619001	  & 	 55538.59 $^{+0.01}_{-0.01}$ & 	 -2967  &  $<$85.48  \\
        Swift-XRT  & 83377001	  & 	 56830.70 $^{+0.01}_{-0.01}$ & 	 -1675  &  $<$262.6  \\
        Swift-XRT  & 83377002	  & 	 58181.14 $^{+1.01}_{-0.01}$ & 	 -324  &  $<$80.28  \\
        Swift-XRT  & 11115001-11  & 	 58535.23 $^{+9.40}_{-20.55}$ & 	 30  &  4.62$^{+1.91}_{-1.58}$  \\
        XMM-Newton${^*}$ & 0831791001   & 	 58543.2  & 	 38   &  4.48$^{+0.77}_{-0.78}$  \\ 
        Swift-XRT  & 11115012-14	  & 	 58561.71 $^{+0.27}_{-0.27}$ & 	 56  &  $<$9.45  \\
        Swift-XRT  & 11115015-22	  & 	 58581.48 $^{+18.86}_{-9.50}$ & 	 76  &  $<$6.07  \\
        XMM-Newton${^*}$ & 0831791101  & 	 58589.0  & 	 84   &  1.24$^{+0.40}_{-0.38}$  \\ 
        CXO-ACIS-S & 22182    & 	 58591.07  & 	 85   &  2.01$^{+0.98}_{-0.56}$  \\
        Swift-XRT  & 11115024-25	  & 	 58618.17 $^{+3.30}_{-0.17}$ & 	 113  &  $<$25.45  \\
        CXO-ACIS-S & 22183    & 	 58643.59  & 	 138  &  3.18$^{+0.77}_{-1.11}$  \\
        Swift-XRT  & 11115026-27	  & 	 58646.39 $^{+4.68}_{-1.37}$ & 	 141  &  6.47$^{+4.39}_{-3.13}$  \\
        Swift-XRT  & 11115028-33	  & 	 58680.73 $^{+10.25}_{-15.18}$ & 	 175  &  4.51$^{+2.54}_{-1.92}$  \\
        Swift-XRT  & 11115034-40	  & 	 58730.84 $^{+17.78}_{-12.43}$ & 	 225  &  4.9$^{+2.24}_{-1.79}$  \\
        Swift-XRT  & 11115041-45	  & 	 58802.80 $^{+5.11}_{-23.69}$ & 	 297  &  $<$12.12  \\
        Swift-XRT  & 11115046-49	  & 	 58829.49 $^{+0.01}_{-7.57}$ & 	 324  &  $<$16.17  \\
        Swift-XRT  & 11115050-54	  & 	 58840.13 $^{+16.66}_{-0.00}$ & 	 335  &  $<$10.6  \\
        Swift-XRT  & 11115055-56	  & 	 58898.48 $^{+7.70}_{-0.07}$ & 	 393  &  $<$26.76  \\
        Swift-XRT  & 11115057-63	  & 	 58940.59 $^{+20.75}_{-18.61}$ & 	 435  &  $<$7.48  \\
        Swift-XRT  & 11115064-65	  & 	 58971.52 $^{+4.25}_{-0.00}$ & 	 466  &  $<$17.11  \\
        Swift-XRT  & 11115066-67	  & 	 59177.46 $^{+6.18}_{-0.00}$ & 	 672  &  $<$54.95  \\
        Swift-XRT  & 11115069	  & 	 59193.66 $^{+0.01}_{-0.00}$ & 	 688  &  $<$76.33  \\
        Swift-XRT  & 11115072	  & 	 59206.46 $^{+1.34}_{-0.00}$ & 	 701  &  $<$35.08  \\
        Swift-XRT  & 11115073	  & 	 59219.41 $^{+0.00}_{-0.00}$ & 	 714  &  $<$696.02  \\
        Swift-XRT  & 11115075-78	  & 	 59247.70 $^{+0.27}_{-14.01}$ & 	 742  &  $<$27.09  \\
        Swift-XRT  & 96581001	  & 	 59811.58 $^{+0.40}_{-0.40}$ & 	 1306  &  $<$27.1  \\
        Swift-XRT  & 96581002	  & 	 59831.46 $^{+0.53}_{-0.01}$ & 	 1326  &  $<$37.73  \\
        Swift-XRT  & 11115079-80	  & 	 60114.40 $^{+3.57}_{-0.01}$ & 	 1609  &  $<$50.62  \\
            \hline 
    \end{tabular}
    \begin{threeparttable}
    \begin{tablenotes}
      \small
      \item * X-ray luminosity measurements  are from grouped EPIC-PN, EPIC-MOS1, and EPIC-MOS2 data, exact values are obtained from \citet{holoien_2019}
    \end{tablenotes}
    \end{threeparttable}
\end{table*}
\subsection{X-rays: Chandra X-ray Observatory}
\label{subsec:ChandraDataAnalysis}
We obtained two epochs of deep X-ray observations of ASASSN-19bt with the Chandra X-ray Observatory (CXO) on 2019-04-17 and 2019-06-09 (Obs IDs 22182 and 22183, PI: K. Alexander, exposure time of 9.98\,ks  for each ObsID), which is $\delta t=$ 85\,days and  $\delta t=$ 138\,days after the optical discovery of the TDE. The data were reduced with the   {\tt CIAO} software package (v4.15) applying standard ACIS data filtering. An X-ray source is blindly detected with {\tt wavdetect} with high statistical confidence $>4\,\sigma$ (Gaussian equivalent) at coordinates consistent with the optical and radio location of the TDE in each of the two CXO observations. The net 0.5-8 keV count-rates are $(8.7\pm3.0)\times 10^{-4}\,\rm{c\,s^{-1}}$ and  $(10.8\pm3.3)\times 10^{-4}\,\rm{c\,s^{-1}}$, respectively. We find no statistical evidence for temporal variability of the source in the CXO dataset. 

For each observation, we extracted a spectrum with {\tt specextract} using a $1.5\arcsec$ source region and a source-free background region. 
Due to the low-number statistics, we cannot statistically constrain the spectral shape and both thermal and non-thermal models are statistically acceptable. Adopting an absorbed simple power-law model ({\tt tbabs*ztbabs*pow} within  {\tt Xspec}), we find best-fitting photon index values in the range $\Gamma=1.6-1.7$ and no evidence for intrinsic absorption. The Galactic neutral hydrogen column density in the direction of ASASSN-19bt is  $N_{H,MW}=7.1\times 10^{20}\,\rm{cm^{-2}}$ \citep{Kalberla05}. We find no statistical evidence for spectral evolution. A joint fit of the two CXO epochs leads to the following best fitting parameters: $\Gamma= 1.64^{+0.57}_{-0.50}$, $N_{H,int}< 3.7\times 10^{22}\,\rm{cm^{-2}}$, where the uncertainties are provided at the $1\,\sigma$ (Gaussian equivalent) confidence level (c.l.) and the upper limit at the $3\,\sigma$ c.l.. These spectral parameters are consistent with those constrained from XMM observations in \citet{holoien_2019}. The  unabsorbed 0.3--10 keV fluxes and luminosities for all X-ray observations used in this work are reported in Table \ref{tab:xray}.

\subsection{X-rays: Swift-XRT}
\label{subsec:SwiftDataAnalysis} 

ASASSN-19bt was monitored with the X-ray Telescope \citep[XRT,][]{Burrows05} onboard the Neil Gehrels  Swift Observatory \citep{Gehrels04}, between 2019-01-31 and 2023-08-15 ($\delta t= 9-1666$ days). We analyzed all available \emph{Swift}-XRT data using HEASoft v.6.32.1 and corresponding calibration files. We extracted a 0.3-10 keV count-rate light-curve using the online automated tools released by the Swift-XRT team \citep{Evans09} \footnote{\href{https://www.swift.ac.uk/user\_objects/}{https://www.swift.ac.uk/user\_objects/}} and custom scripts \citep{Margutti13}. 
A targeted search for X-ray emission from the TDE leads to the detection of a weak but persistent X-ray source with approximately constant count-rate at $\delta t<250$\,days. For the count-to-flux calibration we adopt the best-fitting spectral parameters derived from the joint fit of the two CXO epochs of \S\ref{subsec:ChandraDataAnalysis}, which leads to flux levels comparable to the CXO and XMM inferences. No X-ray source is detected by \emph{Swift}-XRT observations acquired at $\delta t>250$\,days. However, these late-time flux limits are not deep enough to constrain the fading of the source. A persistent X-ray source with $L_x\sim$ a few $10^{40}\,\rm{erg\,s^{-1}}$ cannot be ruled out, which makes the X-rays from ASASSN-19bt among the least luminous ever detected from an optically-selected TDE (Fig. \ref{fig:lit_lc}).  

We note that analyzing data serendipitously acquired by \emph{Swift}-XRT  years before the TDE in the same way leads to the potential identification of a source of X-ray emission in 2009, as previously reported by \citet{holoien_2019}. The source is not blindly detected, but has a targeted detection significance of $3\,\sigma$. If real and attributed to star formation in the host galaxy, the source flux would require a star formation rate of $8.9 ^{+15.2}_{-6.4}$ $M_{\odot}$ yr$^{-1}$ \citep{Riccio_2023}, which is higher than the rates inferred from the archival spectra and radio detections (\S\ref{sec:archival_radio}). Therefore, this marginal detection may indicate prior activity of the central SMBH. We further discuss the X-ray emission from ASASSN-19bt in \S \ref{sec:xray_disc}.


\begin{figure*}[!ht]
    \centering
    \includegraphics[width = \textwidth]{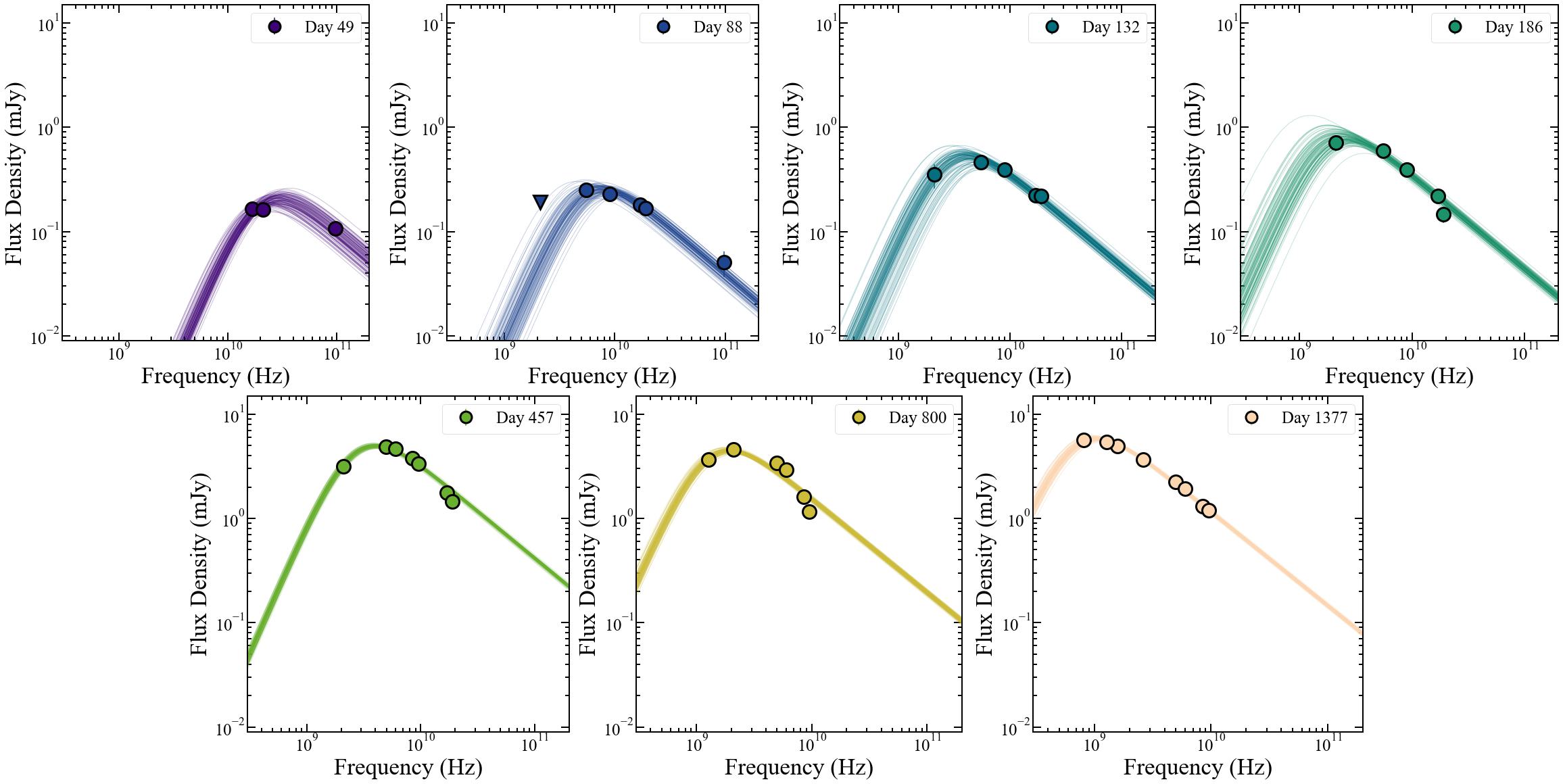}
    \caption{Radio spectral energy distribution fits for the combined ATCA, ALMA, and MeerKAT data obtained by subtracting the modeled quiescent emission component. The solid lines indicate a representative sample of SED fits from the MCMC modeling. The host subtracted SEDs are characteristic of a synchrotron self-absorbed spectrum, with a spectral slope of $F_{\nu}\propto\nu^{5/2}$ below the peak and $F_{\nu}\propto\nu^{(1-p)/2}$ above the peak. The evolution of the SED is atypical of the synchrotron emission expected from an expanding outflow as the peak frequency does not monotonically evolve to lower frequencies over time.}
    \label{fig:1}
\end{figure*}

\section{Synchrotron Emission Modeling}\label{sec:modeling}

We model the radio SEDs as synchrotron emission generated from outflowing material expanding into an external medium. During this process, the blast wave generated by the outflow accelerates the electrons to a power-law distribution of energies, $N(\gamma) \propto \gamma^{-p}$ for $\gamma \geq \gamma_{m,0}$, immediately behind the shock; here, $\gamma$ is the electron Lorentz factor, $\gamma_{m,0}$ is the minimum Lorentz factor of the shocked electrons, and $p$ is the power law index of the energy distribution. The shape of the synchrotron spectrum is in general a multiple broken power-law, described by its break frequencies and an overall normalization factor (e.g., \citealt{Granot_2002}).

For ASASSN-19bt, we assume that the spectral break frequencies follow $\nu_m < \nu_a < \nu_c$, where $\nu_m$ is the typical synchrotron frequency of the minimal electron energy in the power law, $\nu_a$ is the synchrotron self-absorption frequency, and $\nu_c$ is the synchrotron cooling frequency. For the energies and high density environments typical of most TDEs, we expect the SED peak to be located at the synchrotron self-absorption frequency and the other break frequencies to be located outside of the ATCA frequency range \citep[e.g.,][]{alexander_2016, at2019dsg, AT2019azh}. We therefore initially fit our data with a singly broken power law, where the peak frequency is the self-absorption frequency $\nu_a$ with $F_\nu \propto \nu^{5/2}$ below the peak, and $F_\nu \propto \nu^{(1-p)/2}$ above the peak. While the peak frequency is typically associated with $\nu_a$ for non-relativistic sources, the peak is often located at $\nu_m$ for relativistic sources at early times \citep{Eftekhari, Andreoni_2022}. We note that instead considering the peak as $\nu_m$ in our relativistic SED modeling (\S \ref{sec:jet}) wouldn't change the inferred physical parameters significantly.

We group our radio observations into 7 epochs, subtract our model for the quiescent host emission (\S\ref{sec:archival_radio}), and fit the remaining transient component of each SED with our synchrotron model (see Appendix \ref{appendix:sed} for fits to the total flux densities). Although we use the analytic SED shape from the \citet{Granot_2002} model for synchrotron emission from gamma-ray burst (GRB) afterglows for the regime where $ \nu_m \ll \nu_a $, we note that we fit each SED independently, without requiring any specific dynamic evolution of the emission. 
 The model SED is expressed as:

\begin{equation}
    \centering
    F_\nu = F_{\nu,\rm ext}  \left[ \left( \frac{\nu}{\nu_b} \right)^{-s \beta_1}+ \left( \frac{\nu}{\nu_b} \right)^{-s\beta_2} \right]^{-1/s}
    \label{eq:1}
\end{equation} 
where $\beta_1 = {5/2}$ and $\beta_2 = {(1-p)/2}$ are the spectral slopes below and above the break, $s=1.25 - 0.18p$ describes the sharpness of the peak, and $\nu_b = \nu_a$ is the self-absorption frequency. We use the \cite{Granot_2002} analytical expression for $s$ appropriate for an ambient density profile described by $\rho\propto r^{-2}$. This scenario most closely approximates the expected circumnuclear density profile around TDE SMBHs \citep{alexander2020}.

We employed the Markov Chain Monte Carlo (MCMC) module \verb'emcee' \citep{emcee} in Python to determine the optimal model parameters, assuming a Gaussian likelihood for all the parameters. We adopt uniform priors on the model parameters $p$, $F_{\nu,\rm ext}$, and $\nu_a$, where $p \in [2,4]$, log($F_{\nu,\rm ext}/ \rm{mJy}$) $\in [-4,2]$, and log($\nu_a/\rm{Hz}$) $ \in [6,11]$. In the initial modeling stage, we fit for $F_{\nu,\rm ext}$, $\nu_a$, and $p$ in each epoch. Our analysis revealed no significant indication of time-dependent variations in the value of $p$. Therefore, we adopted a weighted average of $p = 2.80 \pm 0.02 $ for the transient component ($p =  2.75 \pm 0.02$ for the total flux density fits). In our subsequent analysis, we set $p$ to be constant at these values and only fit for $F_{\nu,\rm ext}$ and $\nu_a$. We sampled the posterior distributions for $F_{\nu,\rm ext}$ and $\nu_a$ with 100 MCMC chains using the sampler EnsembleSampler from \verb"emcee". To ensure the convergence of the samples, we ran each chain for 2,000 steps and discarded the initial 1,000 steps. We report the best fit transient SED parameters in Table \ref{tab:2} and Figure \ref{fig:1} shows the resulting transient SED fits (for fits to the total flux density, see Figure \ref{fig:1_orig} in Appendix \ref{appendix:sed}).

\begin{figure}[!ht]
    \centering
    \includegraphics[width = 0.45\textwidth]{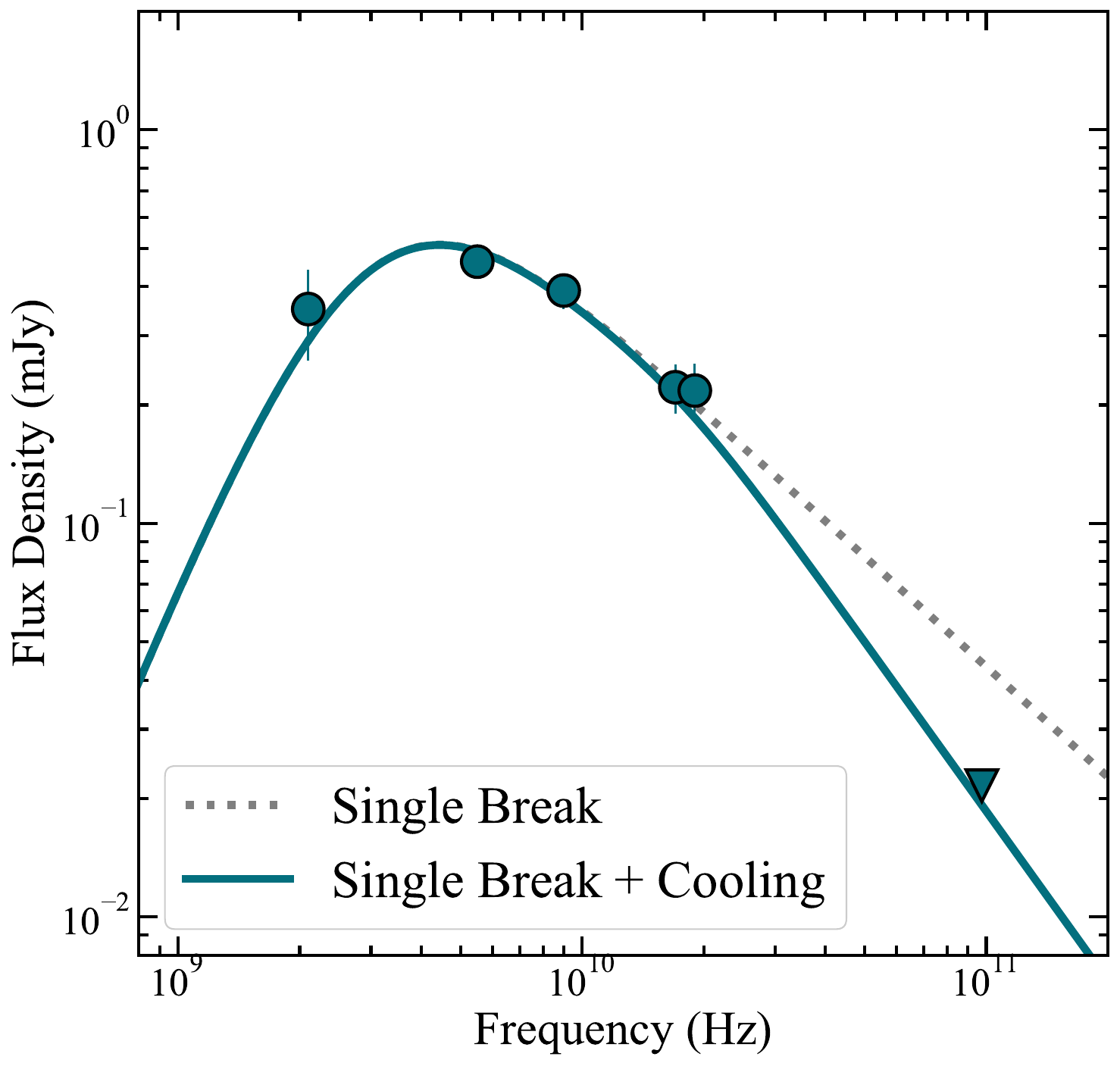}
    \caption{ATCA data along with a non detection from ALMA at day $\langle \delta t \rangle \approx 132$ days. The dashed gray line indicates the model synchrotron self-absorbed spectrum with only one break frequency at $\nu_b \approx \nu_a$. The teal line shows the model spectrum that incorporates synchrotron cooling which finds $\nu_c \approx 19$ GHz leading to $\epsilon_B\approx0.2$ for a spherical Newtonian outflow and $3 \times 10^{-4} < \epsilon_B < 0.005$ for an off-axis jet depending on the viewing angle.}
    \label{fig:3}
\end{figure}

\begin{table}[!h]
\centering
    \begin{tabular}{lccc}
    \hline\hline
        Date & $\langle \delta t \rangle$ & $\nu_p$ & $F_p$ \\%
        (UTC) & (days) & (GHz) & (mJy) \\%
        \hline
        2019 Mar 12  &  49 & 28.56 $\pm$ 3.34 & 0.20 $\pm$ 0.02 \\
        2019 Apr 11/13/19  &  88 & 7.79 $\pm$ 1.22 & 0.25 $\pm$ 0.02 \\
        2019 Jun 3  &  132 & 4.27 $\pm$ 0.64 & 0.53 $\pm$ 0.06 \\
        2019 Sep 27  &  186 & 2.53 $\pm$ 0.41 & 0.79 $\pm$ 0.11 \\
        2020 Apr 23  &  457 & 4.03 $\pm$ 0.12 & 4.96 $\pm$ 0.14 \\
        2021 Mar 25/Apr 10/19  &  800 & 1.95 $\pm$ 0.09 & 4.46 $\pm$ 0.15 \\
        2022 Sep 30/2023 Jan 25  &  1377 & 1.06 $\pm$ 0.05 & 5.78 $\pm$ 0.20 \\
    \hline \hline
    \end{tabular}

    \caption{Spectral energy distribution parameters from the synchrotron emission model. We fit only the transient component of the radio flux densities.    \label{tab:2}
}
\end{table} 

\subsection{Synchrotron Cooling}\label{sec:cool}

On 2019 June 20, we obtained an upper limit on the flux density at 97.5 GHz that is inconsistent with our single-break synchrotron SED model fit to the contemporaneous observations at $\lesssim20$ GHz. This discrepancy may be due to the presence of a synchrotron cooling break at a frequency of $\nu_c$, given by \citep{Sari_1998}:

\begin{equation}
    \nu(\gamma_c) = \frac{q_e B}{2 \pi m_e c} \Gamma \gamma_c^2, 
    \label{eq:2}
\end{equation}
where $\gamma_c$ is the critical Lorentz factor required for synchrotron cooling, defined as $\gamma_c = 6\pi m_e c / \sigma_T B^2 \Gamma t$, $\Gamma$ is the bulk Lorentz factor of the shocked material, and $t$ refers to the time since the launch of the outflow in the frame of the observer. We conservatively assume that the radio outflow was launched at $\delta t=0$ days.\footnote{We note that while the radio outflow was not necessarily launched exactly at $\delta t =0$ days, the launch date must be before the date of our first ALMA detection at $\delta t\approx40$ days. This allowed range accounts for a very small fractional difference in time when considering the later epochs, so this assumption has minimal impact on our results.} 

In Figure \ref{fig:3}, we show the single break SED for the data from day 132, along with a model SED that describes the single break plus synchrotron cooling. The SED model including the cooling break is given by \citep{Granot_2002}: 

\begin{equation}
    \centering
    F_{\nu_c} = F_{\nu}  \left[1+\left( \frac{\nu}{\nu_c} \right)^{s(\beta_2-\beta_3)}\right]^{-1/s},
\end{equation} 
where $F_\nu$ is the model described by equation (\ref{eq:1}), $\beta_3 = {-p/2}$ is the spectral slope above the cooling break, and $s=10$ describes the smoothing \footnote{We note that this smoothing term is much steeper than the suggested value in \citet{Granot_2002}, which was derived in the context of GRB afterglows. Instead, we adopt the smoothing parameter used in \citet{at2019dsg} for the TDE AT2019dsg, which was motivated by the observed sharpness of the break.} describes the smoothing. Our best-fit synchrotron cooling model suggests that $\nu_c \approx 19 \ \rm GHz$. Due to the $B$ dependence of $\nu_c$, measuring the location of the cooling break leads to an estimate of $\epsilon_B$, the fraction of post-shock energy in magnetic fields. We outline our expressions for $B$ explicitly in \S \ref{sec:analysis}. 
The observed $\nu_c$ is consistent with $\epsilon_B \approx 0.2$ under the assumption that the outflow is spherical and non-relativistic. In the off-axis jet scenario, the implied magnetic field strength $B$ becomes a function of the observer viewing angle. We find $\epsilon_B$ ranges from $\sim 3 \times10^{-4}$ to $\sim 0.005$ depending on the viewing angle (for exact values see Table \ref{tab:eb} in Appendix \ref{appendix:sed}). 

\begin{figure}[t]
    \centering
    \includegraphics[width = 0.45\textwidth]{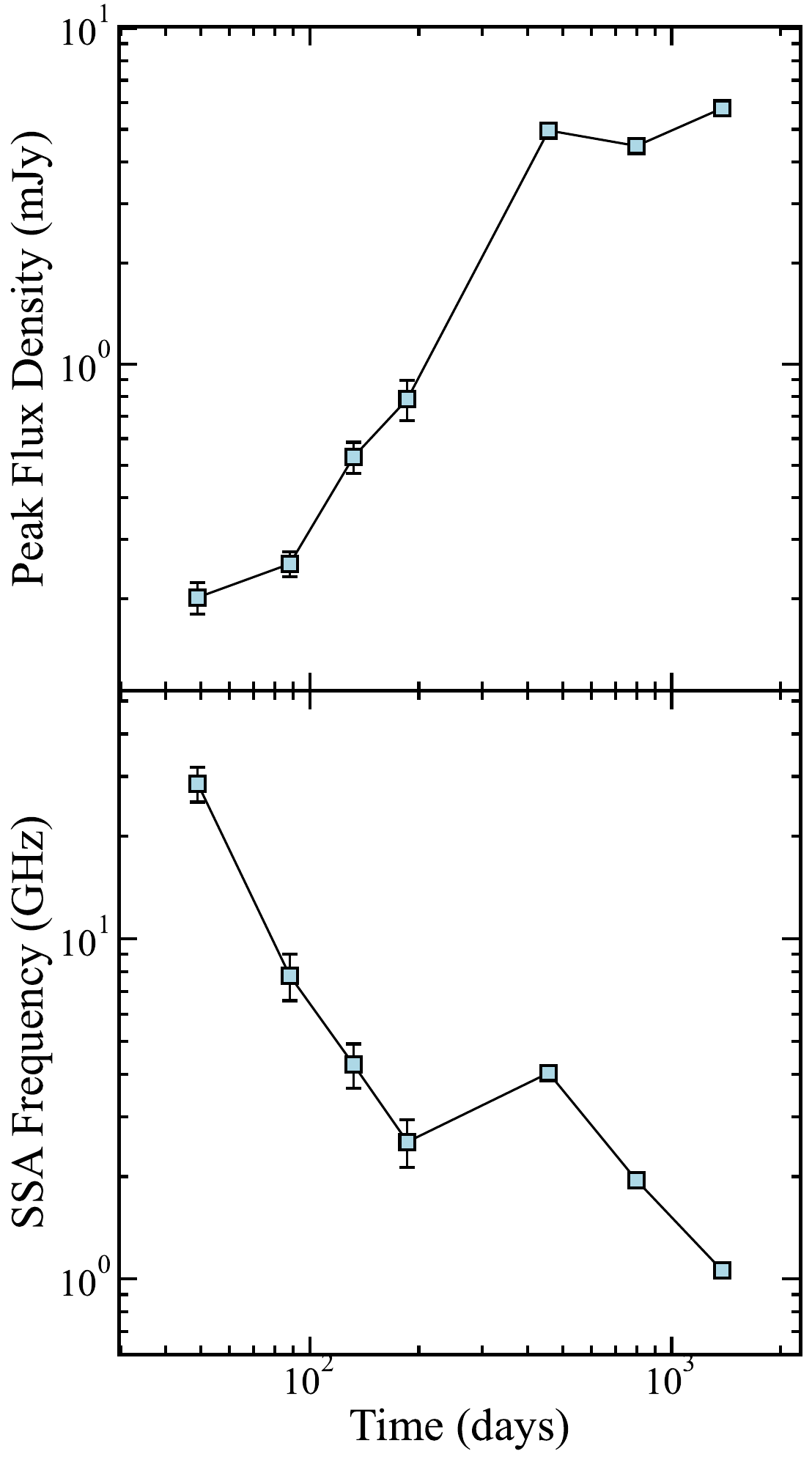}
    \caption{(\textit{Top}) The evolution of the observed peak flux density derived from the SED modeling. The peak flux density increases as a power law at early times then remains approximately constant. (\textit{Bottom}) The evolution of the synchrotron self-absorption (SSA) frequency ($\nu_a$) derived from the SED modeling. $\nu_a$ decays as a power law at early and late times, however the late time evolution is disjoint from the earlier epochs.}
    \label{fig:2}
\end{figure}

\begin{table*}[ht]
    \centering
    \begin{tabular}{ccccccccc}
    \hline\hline
        Geometry & $\langle \delta t \rangle$ & log($R$) & log($E$) & log($E_\text{wide}$) & log($B$) & log($N_e$) & log($n_\text{ext}$) & $\beta$ \\ 
         & ($\text{days})$ & $(\text{cm})$ & $(\text{erg})$ & $(\text{erg})$ & $(\text{G})$ &  & ($\text{cm}^{-3})$ &  \\ 
        \hline
        Spherical & 49 & 15.30 $^{+ 0.05 }_{- 0.04 }$ & 46.40 $^{+ 0.05 }_{- 0.04 }$ & - & 0.86 $^{+ 0.04 }_{- 0.05 }$ & 50.57 $^{+ 0.05 }_{- 0.04 }$ & 4.01 $^{+ 0.09 }_{- 0.09 }$ & 0.016 $^{+ 0.002 }_{- 0.001 }$  \\ 
        $f_A=1$ & 88 & 15.91 $^{+ 0.09 }_{- 0.06 }$ & 47.08 $^{+ 0.12 }_{- 0.09 }$ & - & 0.29 $^{+ 0.07 }_{- 0.07 }$ & 51.26 $^{+ 0.12 }_{- 0.09 }$ & 2.85 $^{+ 0.15 }_{- 0.13 }$ & 0.036 $^{+ 0.008 }_{- 0.005 }$  \\ 
        $f_V=0.36$ & 132 & 16.32 $^{+ 0.08 }_{- 0.07 }$ & 47.72 $^{+ 0.11 }_{- 0.10 }$ & - & -0.01 $^{+ 0.06 }_{- 0.06 }$ & 51.89 $^{+ 0.11 }_{- 0.10 }$ & 2.26 $^{+ 0.14 }_{- 0.11 }$ & 0.062 $^{+ 0.011 }_{- 0.010 }$  \\ 
        $\epsilon_e = 0.1$ & 186 & 16.64 $^{+ 0.11 }_{- 0.08 }$ & 48.16 $^{+ 0.16 }_{- 0.11 }$ & - & -0.26 $^{+ 0.07 }_{- 0.07 }$ & 52.34 $^{+ 0.16 }_{- 0.11 }$ & 1.77 $^{+ 0.15 }_{- 0.12 }$ & 0.090 $^{+ 0.023 }_{- 0.016 }$  \\ 
        $\epsilon_B = 0.2$ & 457 & 16.81 $^{+ 0.02 }_{- 0.02 }$ & 48.92 $^{+ 0.02 }_{- 0.02 }$ & - & -0.14 $^{+ 0.02 }_{- 0.01 }$ & 53.09 $^{+ 0.02 }_{- 0.02 }$ & 2.01 $^{+ 0.03 }_{- 0.03 }$ & 0.055 $^{+ 0.002 }_{- 0.002 }$  \\ 
        & 800 & 17.11 $^{+ 0.02 }_{- 0.03 }$ & 49.18 $^{+ 0.03 }_{- 0.03 }$ & - & -0.45 $^{+ 0.02 }_{- 0.02 }$ & 53.35 $^{+ 0.03 }_{- 0.03 }$ & 1.38 $^{+ 0.05 }_{- 0.04 }$ & 0.062 $^{+ 0.003 }_{- 0.004 }$  \\ 
        & 1377 & 17.42 $^{+ 0.03 }_{- 0.02 }$ & 49.57 $^{+ 0.04 }_{- 0.03 }$ & - & -0.73 $^{+ 0.02 }_{- 0.03 }$ & 53.75 $^{+ 0.04 }_{- 0.03 }$ & 0.83 $^{+ 0.04 }_{- 0.05 }$ & 0.074 $^{+ 0.006 }_{- 0.004 }$  \\ 
        \hline\hline
        
        Geometry & $\langle \delta t \rangle$ & log($R$) & log($E$) & log($E_\text{wide}$) & log($B$) & log($N_e$) & log($n_{ext}$) & $\Gamma$ \\ 
        & ($\text{days})$ & $(\text{cm})$ & $(\text{erg})$ & $(\text{erg})$ & $(\text{G})$ &  & ($\text{cm}^{-3})$ &  \\ 
        \hline
        Jet & 49 & 17.28 $^{+ 0.01 }_{- 0.01 }$ & 50.48 $^{+ 0.07 }_{- 0.07 }$ & 52.21 $^{+ 0.11 }_{- 0.12 }$ & 0.54 $^{+ 0.05 }_{- 0.07 }$ & 52.83 $^{+ 0.05 }_{- 0.04 }$ & 3.31 $^{+ 0.09 }_{- 0.10 }$ & 51.2 $^{+ 3.3 }_{- 3.7 }$  \\ 
        $\theta_\text{obs}$=1.05 & 88 & 17.54 $^{+ 0.01 }_{- 0.01 }$ & 50.35 $^{+ 0.09 }_{- 0.09 }$ & 51.56 $^{+ 0.22 }_{- 0.17 }$ & -0.17 $^{+ 0.10 }_{- 0.09 }$ & 52.86 $^{+ 0.05 }_{- 0.06 }$ & 2.05 $^{+ 0.18 }_{- 0.14 }$ & 28.3 $^{+ 3.4 }_{- 3.7 }$  \\ 
        $f_A = 1$ & 132 & 17.71 $^{+ 0.01 }_{- 0.01 }$ & 50.46 $^{+ 0.07 }_{- 0.06 }$ & 51.31 $^{+ 0.22 }_{- 0.14 }$ & -0.56 $^{+ 0.10 }_{- 0.08 }$ & 53.04 $^{+ 0.03 }_{- 0.04 }$ & 1.35 $^{+ 0.18 }_{- 0.12 }$ & 19.0 $^{+ 2.7 }_{- 2.1 }$  \\ 
        $f_V = 1$ & 186 & 17.86 $^{+ 0.01 }_{- 0.01 }$ & 50.54 $^{+ 0.08 }_{- 0.08 }$ & 51.15 $^{+ 0.23 }_{- 0.17 }$ & -0.86 $^{+ 0.10 }_{- 0.10 }$ & 53.17 $^{+ 0.04 }_{- 0.05 }$ & 0.80 $^{+ 0.19 }_{- 0.15 }$ & 14.4 $^{+ 2.2 }_{- 2.2 }$  \\ 
        $\epsilon_e = 0.1$ & 457 & 18.25 $^{+ 0.01 }_{- 0.01 }$ & 51.78 $^{+ 0.02 }_{- 0.02 }$ & 52.72 $^{+ 0.05 }_{- 0.04 }$ & -0.66 $^{+ 0.02 }_{- 0.02 }$ & 54.24 $^{+ 0.01 }_{- 0.01 }$ & 1.02 $^{+ 0.04 }_{- 0.03 }$ & 20.8 $^{+ 0.6 }_{- 0.5 }$  \\ 
        $\epsilon_B = 0.003$ & 800 & 18.50 $^{+ 0.01 }_{- 0.01 }$ & 51.93 $^{+ 0.02 }_{- 0.02 }$ & 52.79 $^{+ 0.07 }_{- 0.05 }$ & -0.99 $^{+ 0.03 }_{- 0.03 }$ & 54.41 $^{+ 0.01 }_{- 0.01 }$ & 0.39 $^{+ 0.06 }_{- 0.04 }$ & 19.0 $^{+ 0.9 }_{- 0.7 }$  \\ 
         & 1377 & 18.73 $^{+ 0.01 }_{- 0.01 }$ & 52.14 $^{+ 0.02 }_{- 0.03 }$ & 52.89 $^{+ 0.06 }_{- 0.07 }$ & -1.30 $^{+ 0.03 }_{- 0.04 }$ & 54.65 $^{+ 0.02 }_{- 0.02 }$ & -0.20 $^{+ 0.05 }_{- 0.06 }$ & 16.7 $^{+ 0.7 }_{- 0.9 }$  \\ 
        \hline
        Jet & 49 & 17.02 $^{+ 0.01 }_{- 0.01 }$ & 50.24 $^{+ 0.07 }_{- 0.07 }$ & 51.37 $^{+ 0.11 }_{- 0.12 }$ & 0.66 $^{+ 0.05 }_{- 0.07 }$ & 52.70 $^{+ 0.05 }_{- 0.04 }$ & 3.38 $^{+ 0.09 }_{- 0.10 }$ & 25.7 $^{+ 1.6 }_{- 1.9 }$  \\ 
        $\theta_\text{obs}$=1.57 & 88 & 17.27 $^{+ 0.01 }_{- 0.01 }$ & 50.12 $^{+ 0.09 }_{- 0.09 }$ & 50.72 $^{+ 0.22 }_{- 0.17 }$ & -0.04 $^{+ 0.10 }_{- 0.09 }$ & 52.73 $^{+ 0.05 }_{- 0.06 }$ & 2.12 $^{+ 0.18 }_{- 0.14 }$ & 14.2 $^{+ 1.7 }_{- 1.8 }$  \\ 
        $f_A = 1$ & 132 & 17.44 $^{+ 0.01 }_{- 0.01 }$ & 50.23 $^{+ 0.07 }_{- 0.06 }$ & 50.48 $^{+ 0.22 }_{- 0.14 }$ & -0.43 $^{+ 0.10 }_{- 0.08 }$ & 52.90 $^{+ 0.04 }_{- 0.04 }$ & 1.42 $^{+ 0.18 }_{- 0.12 }$ & 9.5 $^{+ 1.3 }_{- 1.1 }$  \\ 
        $f_V = 1$ & 186 & 17.59 $^{+ 0.01 }_{- 0.01 }$ & 50.30 $^{+ 0.08 }_{- 0.08 }$ & 50.31 $^{+ 0.23 }_{- 0.17 }$ & -0.73 $^{+ 0.10 }_{- 0.10 }$ & 53.03 $^{+ 0.04 }_{- 0.05 }$ & 0.87 $^{+ 0.19 }_{- 0.15 }$ & 7.2 $^{+ 1.1 }_{- 1.1 }$  \\ 
        $\epsilon_e = 0.1$ & 457 & 17.98 $^{+ 0.01 }_{- 0.01 }$ & 51.55 $^{+ 0.02 }_{- 0.02 }$ & 51.88 $^{+ 0.05 }_{- 0.04 }$ & -0.53 $^{+ 0.02 }_{- 0.02 }$ & 54.11 $^{+ 0.01 }_{- 0.01 }$ & 1.09 $^{+ 0.04 }_{- 0.03 }$ & 10.4 $^{+ 0.3 }_{- 0.3 }$  \\ 
        $\epsilon_B = 0.005$ & 800 & 18.23 $^{+ 0.01 }_{- 0.01 }$ & 51.69 $^{+ 0.02 }_{- 0.02 }$ & 51.95 $^{+ 0.07 }_{- 0.05 }$ & -0.87 $^{+ 0.03 }_{- 0.03 }$ & 54.28 $^{+ 0.01 }_{- 0.02 }$ & 0.46 $^{+ 0.06 }_{- 0.05 }$ & 9.5 $^{+ 0.5 }_{- 0.3 }$  \\ 
         & 1377 & 18.46 $^{+ 0.01 }_{- 0.01 }$ & 51.91 $^{+ 0.02 }_{- 0.03 }$ & 52.05 $^{+ 0.06 }_{- 0.07 }$ & -1.17 $^{+ 0.03 }_{- 0.04 }$ & 54.51 $^{+ 0.02 }_{- 0.02 }$ & -0.13 $^{+ 0.05 }_{- 0.06 }$ & 8.3 $^{+ 0.4 }_{- 0.4 }$  \\ 
        \hline
        \hline
    \end{tabular}
    \caption{Physical parameters derived from our analysis assuming differing outflow geometries. $\langle \delta t \rangle$ is the time since discovery in units of days, all other parameters are in cgs units.}
    \label{tab:3}
\end{table*}

\begin{figure*}[!ht]
    \centering
    \includegraphics[width = \textwidth]{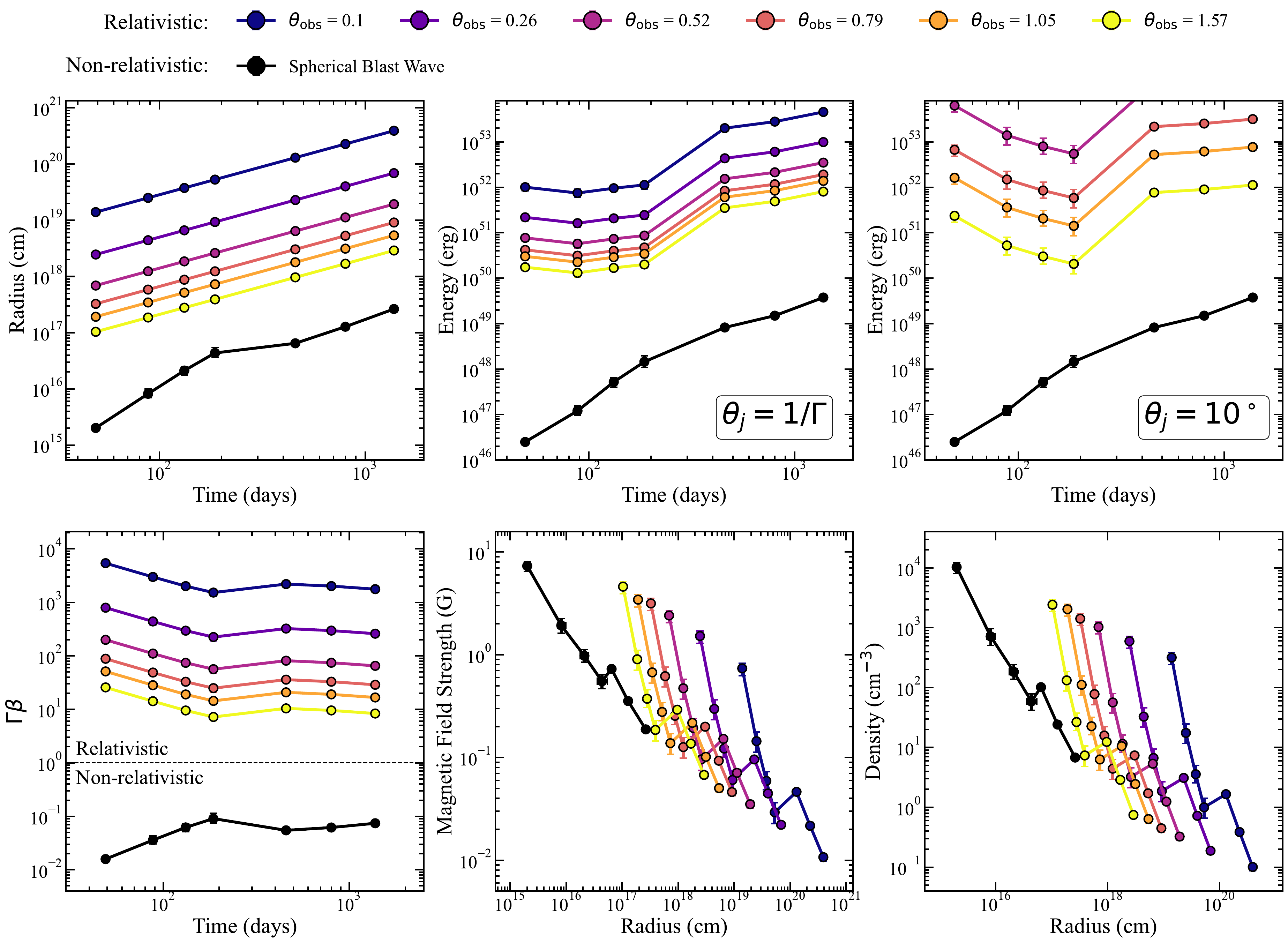}
    \caption{The temporal and radial dependencies of the physical parameters derived from our analysis of the synchrotron model fits. In each panel we show the results for the non-relativistic outflow model presented in \citet{Barniol_Duran_2013} (black circles) and the relativistic solution proposed in \citet{matsumoto} (indigo - yellow circles) for a set of off-axis viewing angles. We show the radius of the emitting region as a function of time (\textit{Top, Left}), the outflow kinetic energy as a function of time for a jet with $\theta_j=1/\Gamma$ (\textit{Top, Middle}),  the outflow kinetic energy as a function of time assuming a fixed jet opening angle $\theta_j=10^\circ$ (\textit{Top, Right}), the velocity evolution of outflow (\textit{Bottom, Left}), the radial profile of the magnetic field (\textit{Bottom, Middle}), and the radial profile of the number density of electrons in the emitting region (\textit{Bottom, Right}). The error bars on the data correspond to 1 standard deviation computed using a Markov Chain Monte Carlo approach. The quantities shown here are summarized in Table \ref{tab:3}.}
    \label{fig:eq_jet}
\end{figure*}

\section{Outflow Modelling}\label{sec:analysis}
The SED fits described in \S 3 allow us to constrain the peak frequency ($\nu_p$) and the peak flux density ($F_p$) in each epoch (Table \ref{tab:2} and Figure \ref{fig:2}). We found that the peak flux density $F_p$ for the transient component increases for the first five epochs, then appears to plateau in the subsequent two epochs. The most dramatic brightening  occurred between $\delta t \sim 186$ days and $\delta t \sim 457$ days post-discovery, with the peak flux density evolving as $t^{2.1}$ and all single-frequency radio light curves increasing by factors of $\sim3-8$. During this time range, we also find that the self-absorption frequency $\nu_a$ increases, which is unusual but not unprecedented when compared to previously seen behavior in other TDEs \citep[e.g.,][]{alexander_2016, Eftekhari, AT2020vwl, AT2018hyz}. We find that $\nu_a$ decays as $t^{-1.4}$ and $t^{-1.3}$ in the epochs preceding and succeeding the brightening event that occurred between days $\delta t \sim 186$ and $\delta t \sim 457$. 
To model the atypical evolution of the SED, we examined two scenarios: (1) the radio emission is caused by a non-relativistic spherical outflow, or (2) the emission is due to a relativistic outflow in the form of a collimated jet viewed off-axis from the line of sight. While we initially assume that the complete radio evolution can be explained by a single outflow, we also discuss the possibility of multiple outflows in \S \ref{sec:split_launch}.

\begin{figure*}
    \centering
    \includegraphics[width = \textwidth]{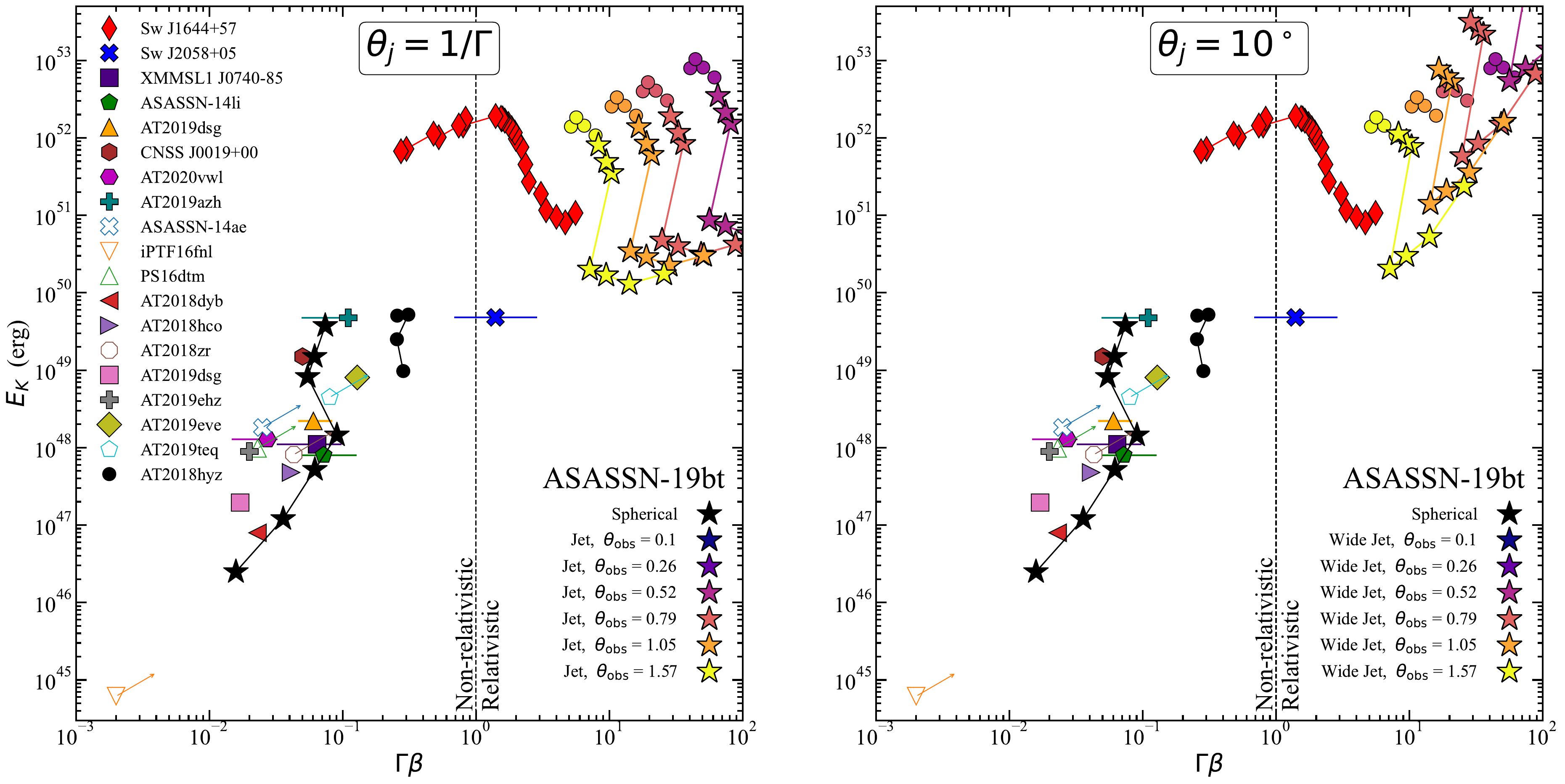}
    \caption{The outflow kinetic energies and velocities inferred from radio emission of known TDEs. The non-relativistic spherical blast wave model is represented in black, while off-axis relativistic jet model is color-coded from indigo to yellow depending on the viewing angle (ASASSN-19bt is denoted by stars, and AT2018hyz is denoted by circles). We show the relativistic model for a jet opening angle of $\theta_j=1/\Gamma$ (\textit{Left}) and a fixed opening angle of $\theta_j=10^\circ$ (\textit{Right}). The data shown for AT2018hyz are from \citet{AT2018hyz} and \citet{matsumoto}. The remaining data are from Sw J1644+57, \citet{Zauderer_2011}, \citet{Eftekhari}, \citet{Cendes_2021}; AT2019dsg, \citealt{at2019dsg,yvette_2023}; ASASSN-14li, \citet{alexander_2016}; AT2020vwl, \citet{AT2020vwl}; AT2019azh, \citet{AT2019azh}; CNSS J0019+00, \citet{CNSS}; Sw J2058+05, \citet{cenko_2012}; all other data are from \citet{yvette_2023}}
    \label{fig:Ek_gamma}
\end{figure*}

\subsection{Non-Relativistic Outflow}\label{outflow:nonrel}
Assuming the observed radio emission is produced by synchrotron radiation generated in the shock between a blast wave and the ambient circumnuclear environment, we can determine the physical properties of the outflow using the procedures outlined in \citet{Barniol_Duran_2013}. In our analysis, we associate $\nu_p$ with $\nu_a$ and assume $\nu_m < \nu_a$. Following \citet{Barniol_Duran_2013}, we define the emitting area $f_A \equiv A/ (\pi R^2/\Gamma^2)$ and the emitting volume  $f_V \equiv V/ (\pi R^3/\Gamma^4)$ and take $\Gamma=1$ for an assumed non-relativistic outflow. To compare to previously studied TDEs from the literature \citep[e.g.,][]{alexander_2016,at2019dsg, AT2018hyz}, we assume that the emitting volume of the outflow is confined to a spherically-symmetric shell of radius $0.1R$, implying $f_A = 1$ and $f_V = \frac{4}{3}(1-0.9^3) \approx 0.36$. With this framework, we can now derive the radius $R$ and energy $E$ using this geometry and the observed values of $F_p$ and $\nu_p$. With $p=2.8$, the expressions for the radius and energy are given by \citep{Barniol_Duran_2013}:

\begin{equation}
    R \approx \   (2.2 \times 10^{17} {\rm cm})\frac{F_{p,{\rm mJy}}^{\frac{44}{93}}d_{L,28}^{\frac{88}{93}} \epsilon^{\frac{1}{17}}\xi^{\frac{1}{93}} 4^{\frac{1}{93}}}{
    \nu_{p,10} (1+z)^{\frac{137}{93}} f_A^{\frac{13}{93}} f_V^{\frac{5}{93}}\gamma_m^\frac{4}{93}}
\end{equation}

\begin{equation}
    E\approx (2.9 \times 10^{50} {\rm erg}) \frac{F_{p,{\rm mJy}}^{\frac{112}{93}}d_{L,28}^{\frac{224}{93}}f_V^{\frac{38}{93}}\xi^{\frac{55}{93}} 4^{\frac{55}{93}} (11+6\epsilon)}{ 
    \nu_{p,10}(1+z)^{\frac{205}{93}}f_A^{\frac{19}{31}} \gamma_m^\frac{44}{93}(17 \epsilon^{\frac{6}{17}})}
\end{equation}
where $F_p$ is scaled to units of mJy, $\nu_p$ to units of 10 GHz, and the luminosity distance $d_L$ in units of $10^{28}$ cm. 
In our analysis we set the minimum Lorentz factor $\gamma_m=2$ and incorporate additional factors of 4 as specified in \citet{Barniol_Duran_2013} for the Newtonian limit. We parameterize deviations from equipartition by defining $\epsilon = (11/6)(\epsilon_B/\epsilon_e)$, where $\epsilon_e$ and $\epsilon_B$ are the fractions of the total energy in electrons and the magnetic field respectively. In our analysis, we adopt the typical value of $\epsilon_e=0.1$ \citep[e.g.,][]{Panaitescu_2002} and fix $\epsilon_B = 0.2$, which is motivated by the observed location of the cooling break at 132 days (\S\ref{sec:cool}). We assume that the kinetic energy is dominated by the protons in the outflow; we thus incorporate the extra energy carried by the hot protons using the correction factor $\xi =1+\epsilon_e^{-1}$ to the total energy \citep{Barniol_Duran_2013}.

With the computed values of $R$, we can also solve for the magnetic field strength $B$, the Lorentz factor $\gamma_e$ of the electrons that radiate at $\nu_p$, and the total number of electrons in the observed region $N_e$ (see equations 14, 15, 16 in \citealt{Barniol_Duran_2013}). The inferred values of $N_e$ and $R$ allow us to estimate the density of electrons in the ambient medium at radius $R$ as $n_\text{ext} = n_e/4$ where $n_e = N_e/V$ is the number density of the electrons in the outflow and the factor of $1/4$ accounts for the shock jump conditions. Here, $V$ is the emitting volume as defined above: a spherical shell with a thickness of $0.1R$ just behind the blast wave. The results of our analysis are shown in Table \ref{tab:3}.

In Figure \ref{fig:eq_jet}, we show the temporal evolution of the emitting region size $R$ and the corresponding minimum equipartition energy $E$. The size of the synchrotron emitting region increases monotonically with time as one would expect for an expanding outflow. Including all epochs, the radius expands at a rate of $R\propto t^{1.3}$, however the four observations in 2019 ($\delta t = 49 - 186$) follow a much steeper trend of $R\propto t^{2.3}$, and from 2020 onward ($\delta t = 457 - 1377$) the radius increases as $R\propto t^{1.3}$. The minimum energy of this region also shows a positive trend in time, increasing more than three orders of magnitude as $E\propto t^{2.2}$ over the full four years of observation. The inferred specific momentum ($\Gamma \beta$) also goes up with time for some portion of the light curve. We also construct a radial profile for the magnetic field strength (see Figure \ref{fig:eq_jet}). We find that the magnetic field decays with increasing distance as $B(r)\propto r^{-0.8}$ at early times and as $B(r)\propto r^{-1}$ at late times. For discussion of the density profile see Section \S \ref{sec:density}.

\subsection{Relativistic Jet from an Arbitrary Viewing Angle}\label{sec:jet}
Given the nearby ($z=0.0262$) location of this TDE, the radio luminosity, evolution timescale, and lack of detected $\gamma$-ray emission suggest that this outflow is not due to a relativistic jet pointed on-axis towards the observer. However, the appearance of a relativistic or collimated outflow can change due to the viewing angle. If a relativistic jet were to be launched off-axis, the emission would become suppressed at early times due to relativistic beaming away from the viewer. As the jet decelerates, the observer may begin seeing emission from inside the beaming cone.

We examined the possibility of a relativistic jet in ASASSN-19bt by employing the 
model outlined in \citet{matsumoto}. This work generalizes \citet{Barniol_Duran_2013}'s analysis of an on-axis relativistic jet in energy equipartition to consider a jet observed from arbitrary viewing angles. Here, we briefly summarize this model and update the equations from \citet{matsumoto} to account for deviations from energy equipartition and the additional kinetic energy carried by hot protons, as we did in \S \ref{outflow:nonrel}. Unless otherwise specified, all variables have the same definitions as in \S \ref{outflow:nonrel}. 

In the relativistic limit, there are two additional degrees of freedom: the jet Lorentz factor $\Gamma$ and the viewing angle $\theta_\text{obs}$ between the observer's line of sight and the source's direction of motion. The energy therefore does not have a global minimum and we require an additional constraint on the system to solve for the physical parameters of the jet as we did above. We obtain this constraint by making an assumption about the outflow launch time, following \citet{matsumoto}. As in \S \ref{sec:cool}, we take the launch time to be $\delta t=0$ days.

Following the procedure outlined in \citet{matsumoto}, we first solved for the radius that minimizes the total energy under a $p=2.8$ power law distribution of electrons. 
%
%
In this model, the observed location of the cooling break implies $3\times 10^{-4} < \epsilon_B < 0.005$ depending on the viewing angle (Table \ref{tab:eb} in Appendix \ref{appendix:sed} gives the precise values). This suggests that the system is not in equipartition; therefore, as in the non-relativistic case, we incorporate additional terms to accommodate this deviation following \citet{Barniol_Duran_2013}. For $p=2.8$, the corresponding radius and energy are:

\begin{equation}
    R \approx \   (2.4 \times 10^{17} {\rm cm})\frac{F_{p,{\rm mJy}}^{\frac{44}{93}}d_{L,28}^{\frac{88}{93}} \Gamma  \epsilon^{\frac{1}{12}}\xi^{\frac{1}{93}}}{
    \nu_{p,10} (1+z)^{\frac{137}{93}} f_A^{\frac{13}{93}} f_V^\frac{5}{93} \gamma_m^\frac{4}{93} \delta_D^\frac{13}{31}}
\end{equation}

\begin{equation}
    E \approx (6.6 \times 10^{50} {\rm erg}) \frac{F_{p,{\rm mJy}}^{\frac{112}{93}}d_{L,28}^{\frac{224}{93}}f_V^{\frac{38}{93}}\xi^{\frac{55}{93}} \Gamma(11+6\epsilon)}{ 
    \nu_{p,10}(1+z)^{\frac{205}{93}}f_A^{\frac{19}{31}} \gamma_m^\frac{44}{93} \delta_D^\frac{81}{31}(17\epsilon^{\frac{5}{12}})}
\end{equation}

\noindent In this case, $\gamma_m = \chi_e (\Gamma-1)$, where $\chi_e = (p-2/p-1)\epsilon_e(m_p/m_e)$, and $\delta_D\equiv1/\Gamma(1-\beta \cos{\theta_\text{obs}})$ is the relativistic Doppler factor for a given source velocity $\beta=\sqrt{1-1/\Gamma^2}$. The additional terms involving $\xi$ and $\epsilon$ account for hot protons and deviations from equipartition with the same definitions as in \S \ref{outflow:nonrel}. We solve for $\Gamma$ in each epoch using the same procedure outlined \citet{matsumoto}.

We initially set the jet area and volume filling factors to be $f_A=f_V=1$. This accounts only for energy within an angle of $1/\Gamma$ from our line of sight. Material at larger angles will contribute negligibly to the observed emission due to relativistic beaming, so this provides the minimum energy required to explain our observations at each epoch. However, a jet with an evolving opening angle of $\theta_j=1/\Gamma$ is not physically likely; while we do eventually expect lateral spreading of the jet as it decelerates, theoretical work suggests that jet spreading is likely not yet important at the large bulk $\Gamma$ values required by our observations \citep{duffell_2018}. We therefore also consider other possible jet geometries in which the jet opening angle remains fixed independent of $\Gamma$. We did not include any narrow jet geometries ($\theta_j < 1/\Gamma$) as the implied jet opening angle became nonphysical due to large bulk $\Gamma$s. We do investigate the effect of a wide jet geometry by accounting for additional energy outside the beamed emission region within an angle of $1/\Gamma$, which contributes negligibly to the observed emission but increases the total energy of the jet. The true collimation of TDE jets is still unknown. AGN jets are typically seen to have $\Gamma \lesssim 10$ with opening angles around $\theta_{j}\sim20^\circ$ whereas GRBs have been seen to launch highly collimated jets with $\Gamma\sim 100$ with opening angles often below $10^\circ$ (\citealt{Goldstein_2016}; \citealt{Pushkarev_2017}; \citealt{Rouco_2023}). 
Therefore, for our analysis, we impose an intermediate jet opening angle of $\theta_{j} = 10^\circ$. For this fixed opening angle, we are always in the regime where $\theta_{j} > 1/\Gamma$ and subsequently correct for the unseen material by multiplying the equipartition energy by a factor of $4\Gamma^2(1 - \cos{\theta_j})$.
The other derived quantities are unaffected, as they depend only on the emission within the observable region (so we can keep $f_A=f_V=1$ throughout).

In Figure \ref{fig:eq_jet}, we show our derived physical parameters for six possible viewing angles: $\theta_\text{obs} = 0.1,0.26,0.52,0.79,1.05,1.57$ (i.e., $6^\circ, 15^\circ, 30^\circ,  45^\circ, 60^\circ$, and $90^\circ$). We find that the solutions become exceedingly extreme for small viewing angles. Solutions for the most on axis case we tested ($\theta_\text{obs} \approx 6^\circ$) resulted in unphysical outflow velocities of $\Gamma>1000$ and energies approaching $10^{54}$ erg. These solutions diverge to higher energies for small $\theta_\text{obs}$ due to the large beaming effects needed to reduce the observed flux towards the observer. We find that our most reasonable solutions appear to be when we consider the relativistic outflow as a highly off-axis collimated jet (the two most off-axis cases are shown in Table \ref{tab:3}). A jet truly spreading with $\theta_j = 1/\Gamma$ would exhibit two periods of roughly constant energy at $\sim 2 \times 10^{50}$ erg from $\sim$ 50 - 190 days and $\sim 7 \times 10^{51}$ erg from $\sim$ 460 - 1380 days. The wide jet geometry with $\theta_j = 10^\circ$ instead implies that the energy decreases by a factor of $\sim10$ during the first 200 days, then remains constant at $E\sim 10^{52} \ \text{erg}$ at later times. All viewing angles yield a solution where the jet propagates at nearly light speed with $\Gamma\sim10-20$ for the most off-axis viewing angles. We find that the jet slows as $\Gamma\propto t^{-1}$. After the radio re-brightening, we find that $\Gamma$ exhibits a weaker decay where $\Gamma\propto t^{-0.2}$. We note that $\Gamma$ is dependent on the inferred launch date of the jet, for discussion of alternative launch dates see Section \S \ref{sec:split_launch}.

\section{Discussion}\label{sec:disc}

Each of the two models discussed in \S \ref{sec:analysis} provides a reasonable fit to the data, but each has some implications that are difficult to explain. Here, we present the implications of these two models and  compare ASASSN-19bt to other TDEs in the literature.

\begin{figure*}[ht!]
    \centering
    \includegraphics[width = \textwidth]{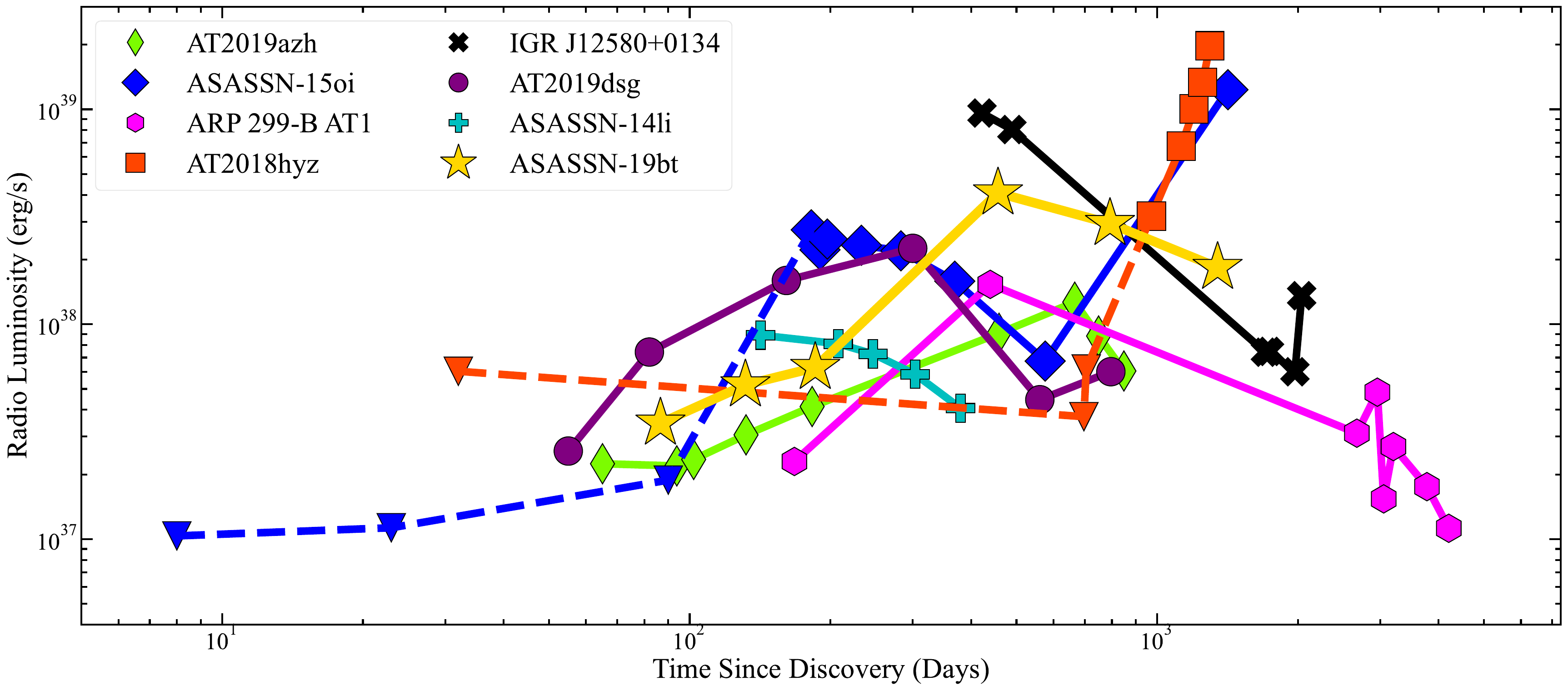}
    \caption{Radio luminosity of ASASSN-19bt at C band ($\approx5.5$ GHz) in context to a selection of previously studied TDEs with radio observations. TDE data are from ASASSN-14li, \citet{alexander_2016}; \citet{AT2019azh}; AT2018hyz, \citet{AT2018hyz}; IGR J12580+0134, \citet{IGR_Lei}, \citet{perlman_2017}; \citet{perlman_2022}; AT2019dsg, \citet{at2019dsg,yvette_2023}; ASASSN-15oi, \citet{horesh_15oi}; ARP 299-B AT1, \citet{ARP_2018}; CNSS J0019+00 \citet{CNSS}.
    }
    \label{fig:lit_lc_radio}
\end{figure*}

Aspects of ASASSN-19bt's radio light curve resemble those of a few other known TDEs (Figure \ref{fig:lit_lc_radio}) as 
many have displayed delayed radio brightening a few years after discovery \citep[e.g.,][]{horesh_15oi, iPTF16fnl, AT2019azh, AT2018hyz, yvette_2023}. At early times, ASASSN-19bt appears most similar to the slow rise seen in AT2019azh \citep{AT2019azh}. AT2019azh also exhibited a late-time flare at a similar timescale to ASASSN-19bt's brightening \citep{sfaradi_19azh}. 
The origin of the radio emission in AT2019azh was difficult to determine, with only the unbound debris stream ruled out as a possibility \citep{AT2019azh}. The late-time evolution and observed luminosity of ASASSN-19bt is also similar to IGR J12580+0134 and ARP 299-B AT1, TDEs with suspected off-axis relativistic jets \citep{IGR_Lei,ARP_2018}. However, its radio evolution is quite different from that of two other TDEs that may harbor powerful jets: AT2018hyz, which may have an off-axis jet and continues to brighten rapidly $>1000$ days post-discovery, and Sw J1644+57, which had an on-axis jet but remained much more luminous than ASASSN-19bt at every phase of its evolution (even at late times when its jet has decelerated and the observed emission should be independent of the viewing angle). We develop these comparisons further below. 

\subsection{Energy and Velocity}\label{sec:ev}

We show the outflow kinetic energy and inferred velocity of ASASSN-19bt in comparison to those of known TDEs in Figure \ref{fig:Ek_gamma}. As mentioned previously, ASASSN-19bt exhibits a dramatic energy increase under the Newtonian model, increasing by 3 orders of magnitude over the period of our observations. This is by far the largest energy increase ever seen in any TDE. 
The off-axis jet scenario also requires the energy to increase with time, with a total energy in the final epoch $\sim50$ times larger than in the first epoch for $\theta_j = 1/\Gamma$ and $\sim5$ times larger for $\theta_j = 10^\circ$.

While energy injection has been inferred for TDEs before (e.g. \citealt{berger_2012} for Sw J1644+57, but see also \citealt{Beniamini_2023}), the much larger energy increase required for ASASSN-19bt is difficult to explain. The energy increase observed in Sw J1644+57 was by a factor $\sim20$ \citep{Eftekhari} and energy increases were also seen in the non-jetted TDEs AT2019dsg (factor $\sim10$;  \citealt{at2019dsg}) and AT2019azh (factor $\sim100$; \citealt{AT2019azh}). Our non-relativistic shell model for ASASSN-19bt indicated $E\propto t^{2.2}$ on average, with a steeper dependence during the first four epochs and a shallow dependence for the remaining three epochs. In the off-axis jet model with $\theta_j =1/\Gamma$, the energy undergoes shallower growth with $E\propto t^{0.1}$ at early times and $E\propto t^{0.7}$ at late times. The energy is declining then flat for the $\theta_j=10^{\circ}$ jet. 

While the jet energy we calculate for ASASSN-19bt for the most off-axis viewing angles is similar to that computed for the jetted TDE Sw J1644+57, our model implies a surprisingly large $\Gamma$ that persists through our last observation at $\delta t\sim1377$ days (Figure \ref{fig:eq_jet}, top right panel). This is similar to AT2018hyz, which was modeled by \cite{matsumoto} as an off-axis jet. When we repeat their analysis under the same assumptions used for ASASSN-19bt we find that modeling AT2018hyz as an off-axis jet requires a slightly higher energy, with slightly lower but still relativistic $\Gamma$ at late times (see Figure \ref{fig:Ek_gamma}). In contrast, Sw J1644+57's jet had decelerated to non-relativistic speeds ($\Gamma\sim1$) by $\delta t\sim700$ days \citep{Eftekhari}. If the relativistic jet model is correct, it would imply that ASASSN-19bt launched the most highly relativistic outflow seen in any TDE to date. It is difficult to understand how such a jet would remain relativistic to such late times in the high-density nuclear environments in which TDEs occur. 
The velocity evolution is equally perplexing in the Newtonian model; while the outflow velocities in the non-relativistic model are consistent with the sample of optically discovered thermal TDEs, ASASSN-19bt's outflow would have to speed up with time (Figure \ref{fig:eq_jet}, bottom left panel), which is difficult to understand physically. In steep density profiles, Newtonian shocks can accelerate; however, this requires a much steeper density profile than we see for ASASSN-19bt \citep{Govreen-Segal_2021}.

\subsection{Circumnuclear Density}\label{sec:density}

\begin{figure*}[ht!]
    \centering
    \includegraphics[width = 0.9\textwidth]{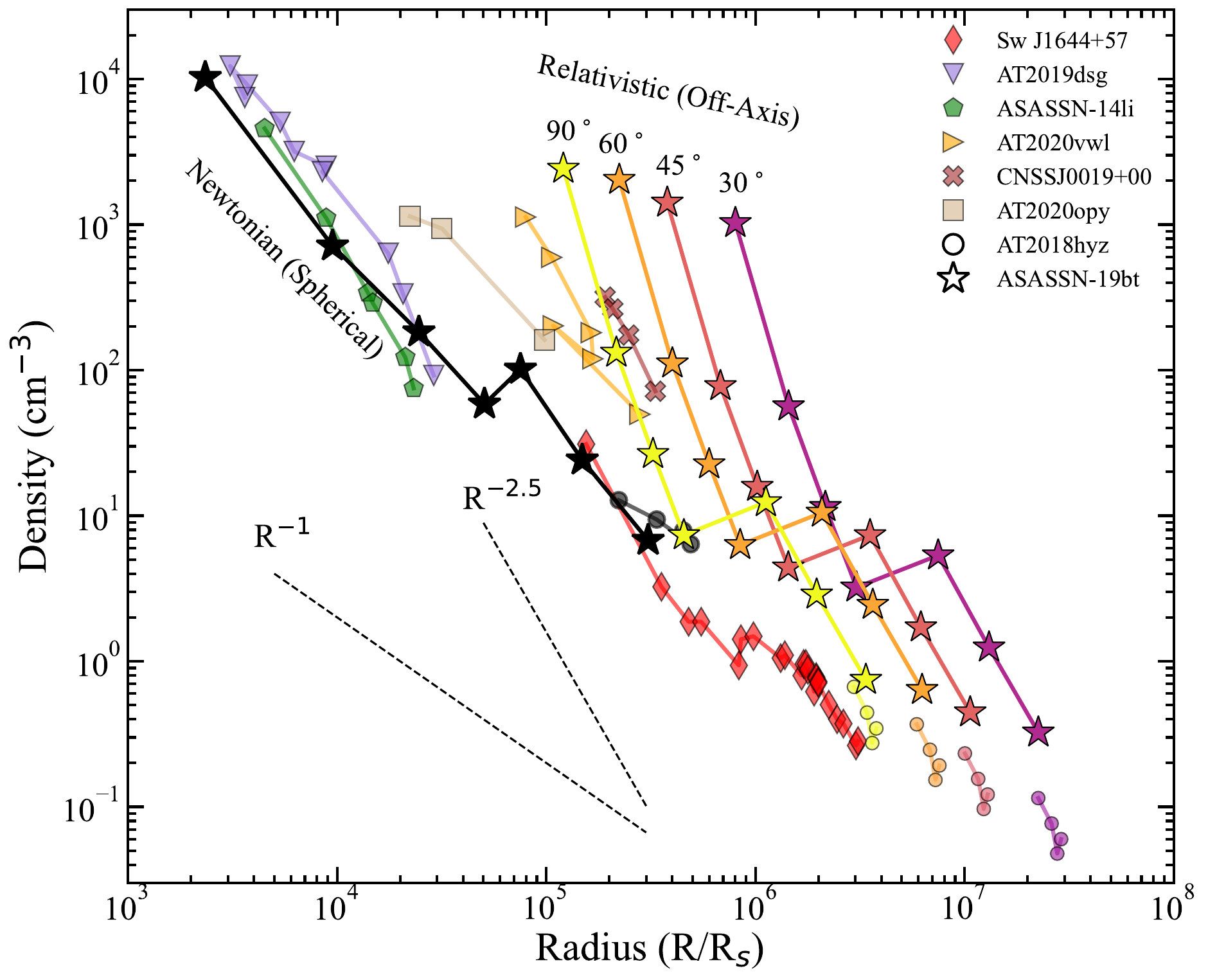}
    \caption{The inferred host galaxy circumnuclear density profile for ASASSN-19bt in context with other TDE host galaxies. To standardize the comparison, we scale the radii by the Schwarzschild radius ($R_s = 2 G M_{\rm BH}/c^2$) of the SMBH at the center of the host galaxy. The non-relativistic spherical blast wave model is represented in black, while off-axis relativistic jet model is color-coded from indigo to yellow depending on the viewing angle (ASASSN-19bt is denoted by stars, and AT2018hyz is denoted by circles). The data shown for AT2018hyz are from \citet{AT2018hyz} and \citet{matsumoto}. The remaining data and assumed SMBH masses are from Sw J1644+57, \citet{Zauderer_2011}, \citet{Eftekhari}, \citet{Cendes_2021}; AT2019dsg, \citet{at2019dsg}; ASASSN-14li, \citet{alexander_2016}; AT2020vwl, \citet{AT2020vwl}; AT2020opy, \citet{at2020opy}; AT2019azh, \citet{AT2019azh}; CNSS J0019+00.}
    \label{fig:density_profile}
\end{figure*}

In Figure \ref{fig:density_profile} we show the inferred circumnuclear density profiles for our ASASSN-19bt models compared to previously studied TDEs with observed radio emission. To standardize the comparison, we scale the radii by the Schwarzschild radius ($R_s = 2 G M_{\rm BH}/c^2$) of the SMBH at the center of each TDE host galaxy and recompute the inferred densities for the thermal TDEs ASASSN-14li, AT2019dsg, CNSS J0019+00, AT2018hyz, AT2020opy, and AT2020vwl using a standard definition for the volume of the emitting region (a spherical 0.1$R$ shell). We use the SMBH masses from Alexander et al.\ 2024 (in prep.) and references therein (particularly \citealt{Yao_2024}). 

The outflow from ASASSN-19bt probes a relatively large fraction of the circumnuclear environment when compared to that of other previously studied TDEs. Both the Newtonian and jet models appear to recover the presence of two separate radial profiles for the magnetic field strength and number density of emitting electrons, corresponding to the pre-brightening and post-brightening portions of the radio light curves. 
We find that the density profile under the non-relativistic model exhibits a $n \propto r^{-3/2}$ slope similar to other thermal TDEs such as AT2020opy and the early epochs of AT2019dsg \citep{at2019dsg, at2020opy}. This is consistent with the classical expectation for spherical Bondi accretion \citep{Bondi_1952}. 
After the blast wave reaches $R\sim 10^{16.5}$ cm, the density increases briefly then continues to decay at a slightly steeper rate $n(r)\propto r^{-1.9}$. This is within the range of density profiles previously observed in TDEs, as a 
$n \propto r^{-2.5}$ profile was measured for ASASSN-14li 
(\citealt{alexander_2016}; and CNSS J0019+00 
\citealt{CNSS}).

In the relativistic model, we estimate the density of the unshocked ambient medium at radius $R$ as $n_\text{ext} = n_e/4\Gamma^2$  where $n_e = N_e/V$ is the number density of emitting electrons behind the blastwave and $V = \pi R^3/\Gamma^4$ is the volume of the emission region \citep{Barniol_Duran_2013}. From this we find that the constructed density profile decays as $n(r)\propto r^{-4.3}$ at early times and as $n(r)\propto r^{-2.6}$ at late times, which is the steepest density profile seen around any TDE to date. We find that the ASASSN-19bt exhibits a similar density profile to that of AT2018hyz if the latter is modeled as an off-axis jet \citep[see][]{matsumoto}, but our observations span a much wider range of circumnuclear radii.

Along with the steepness of the density profile, the off-axis jet models also yield higher densities at large radii compared to other TDEs (for both ASASSN-19bt and AT2018hyz). Assuming that the emitting electrons were swept up by the jet, we can approximate swept up mass as $M_\text{swept}\sim m_\text{p} N_e \left({\theta_j}\right)^2\left(1/\Gamma_\text{latest} \right)^2 \simeq 3 \times 10^{-3} \text{M}_\odot$ for the most off-axis case, assuming a fixed opening angle of $\theta_j = 10^\circ$. As the jet is still relativistic and has not yet decelerated, this implies the total kinetic energy responsible for such a jet is $E_\text{kin} \gtrsim \Gamma M_\text{swept} c^2 \simeq 10^{53} \text{erg}$ (in the more conservative case where $\theta_j = 1/\Gamma_\text{latest} < 10^\circ$, then the required energy is lower by a factor $\sim2$). This energy is among the largest ever inferred for any extragalactic transient; however, it is similar to the energy required by several radio modellings for Swift J1644+57 \citep{matsumoto}. 


\subsection{Multiple Outflows?}{\label{sec:split_launch}}

The models described in \S \ref{sec:analysis} both assume that a single impulsive outflow powers the entire observed radio evolution. However, as previously noted, $F_p$ increases dramatically between $\delta t \sim 186$ days and $\delta t \sim 457$ days. This abrupt radio brightening, together with the presence of a clear discontinuity in the outflow parameters between days $\delta t \sim 186$ and $\delta t \sim 457$ days in both models may imply the presence of two outflows contributing to the radio emission on different timescales. This scenario would make ASASSN-19bt similar to ASASSN-15oi, which showed two distinct emission episodes in its radio light curve \citep{horesh_15oi}. Modeling of ASASSN-15oi's radio emission revealed that the second emission peak likely corresponds to a more energetic outflow launched several hundred days post-disruption (Hajela et al. in prep).

Following \citet{alexander_2016} and \citet{yvette_2023}, we examined the evolution of ASASSN-19bt's emission radius to estimate the outflow launch time for the pre- and post-brightening subsets of the data. For the non-relativistic model outlined in \citet{Barniol_Duran_2013}, if the outflow is assumed to be moving at a constant velocity then we can apply a linear fit to the radius evolution and extrapolate to $R = 0$. However, applying a linear fit to all of the data implies a launch date of $\delta t \approx46$ days, which is inconsistent with our initial radio detection at $\delta t = 40$ days. A linear fit to just the early epochs prior to the re-brightening yields a similarly unphysical launch date of $\delta t \approx52$ days and overall provides a poor fit to the data. If instead we fit the radius evolution in the early epochs with a power law, we find that a much better fit to the data can be obtained if the emitting region is expanding as $R \propto t^{2.3}$. This however suggests a dramatic acceleration of the outflow during the first 186 days, which is difficult to explain (as noted in Section \ref{sec:ev}). Linearly extrapolating the radius evolution for the later epochs ($\delta t > 1$ year) separately would imply a launch date of $\delta t \approx 210$ days, possibly implying that a second outflow launched substantially after the time of disruption powers the latter half of the radio light curve. 
 
For the relativistic solutions outlined in \citet{matsumoto}, a launch date must be assigned in order to solve for the outflow velocity $\Gamma$; for our analysis we set this to be $\delta t = 0$ days. We found that the resulting off-axis jet solutions preferred outflow velocities with $\Gamma \gtrsim 8$. For all viewing angles, $\Gamma$ decreases by a factor $\sim3$ over the first 186 days, then increases slightly at 457 days, and then resumes slowly decreasing. 

Given the suggested launch date for the later epochs in the non-relativistic case, we considered the possibility of two relativistic outflows, where one outflow was launched at $\delta t = 0$ days and the other was launched $\sim210$ days later. This still requires $\Gamma \approx 8$ for the emission at $\delta t > 1$ yr in the most off-axis case and does not fully resolve the energy and density discontinuities.  Although this is perhaps more physically plausible than the accelerating outflow required in the non-relativistic case, it is still surprising that even for the most off-axis viewing angles, $\Gamma$ remains significantly $>1$ even in our last observation at 1377 days.

\subsection{Outflow Mechanism}\label{sec:mechanism}

In order to assess the likelihood of either model, we need to evaluate whether these results are consistent with existing theories describing the origin of radio emission from TDEs. Most TDEs that have received extensive follow-up in the radio are modelled using the Newtonian formalisms outlined in \citet{Barniol_Duran_2013}, therefore, this model is most easily comparable to other TDEs in the literature. The bulk outflow velocity implied by the Newtonian model ($\beta\approx0.05$) matches well with the expected velocities of $\beta\approx0.01-0.1$ for winds due to accretion onto a SMBH \citep[e.g.,][]{Strubbe, Tchekhovskoy}. However, outflows launched with these velocities could also be explained by the unbound debris propagating into the circumnuclear medium or ejecta from a collision-induced outflow (CIO) where the debris fallback stream intersects with itself \citep[e.g.,][]{krolik_2016, bonnerot_2020}. Nonetheless, these non-relativistic outflow scenarios all fail to address the apparent acceleration of the outflow in the first 186 days from $\beta\approx0.016$ to $\beta\approx0.09$. 
Moreover, while increasing energy has previously been found to be expected in some non-relativistic models for TDEs \citep[e.g.,][]{matsumoto_19dsg}, ASASSN-19bt's exceptionally rapid energy increase may be challenging to explain. If the thousandfold increase in the energy is to be attributed to energy injection, it is difficult to conceive of a central engine that remains continuously active for years after the star had been initially disrupted. For example, if we assume that the central engine is only active during the super-Eddington phase, then one would expect the energy injection to stop after approximately 500 days as that was thought to be the duration of the super-Eddington phase for Sw J1644+57 based on a rapid shutoff of the X-ray emission at that time \citep{Mangano_2016,berger_2012}. Other on-axis jetted TDEs, such as AT2022cmc, show even earlier jet shutoff times \citep{eftekhari_2024}. Alternately, energy injection can occur in transients with radially stratified velocity in their ejecta, as previously proposed for some TDEs (e.g.~Sw J1644+57; \citealt{berger_2012}) and other transients (e.g.~GW170817; \citealt{li_2018}). However, the spherical outflow model for ASASSN-19bt requires a larger and more rapid energy increase than either of these cases. 

The off-axis relativistic jet scenario avoids the dramatic energy injection required by the non-relativistic outflow, but is also puzzling. We find that the minimal energy of the outflow appears roughly constant at $\sim 2 \times 10^{50}$ erg at early times, then increasing to $\sim 7 \times 10^{51}$ erg at late times for the most off-axis cases ($\theta_\text{obs} \gtrsim 60^\circ$). For a jet with $\theta_j=10^{\circ}$, the energy decreases by a factor of $\sim10$ during the first 200 days, then remains constant at $E\sim 10^{52} \ \text{erg}$ at later times. It is not clear that this initial energy decrease is physical. Furthermore, as discussed in Section \ref{sec:density}, a higher total kinetic energy is required such that $M_\text{jet} > M_\text{swept}$ to avoid complete deceleration of the jet.

A similar increase in energy was observed in the prototypical jetted TDE Swift J1644+57 on a similar timescale \citep{Eftekhari}. In Swift J1644+57, this energy increase may have also been accompanied by a slight density enhancement, similar to (but less dramatic than) the discontinuity in the density profile seen around ASASSN-19bt that accompanies its energy increase \citep{Eftekhari}. \citet{berger_2012} argued that the energy injection found in Swift J1644+57 could not be explained by further accretion of the fallback stream and instead is likely due to the relativistic jet launching with a wide range of initial Lorentz factors. However, recently, \citet{Beniamini_2023} argued that the observed energy increase can be explained purely by the jet being initially viewed slightly off-axis; as the jet slowed down, the beaming cone opens up to reveal more emission. 
This explanation does not work for ASASSN-19bt as $\Gamma$ increases slightly between the two epochs where the energy jumps. Instead, the discrepancy between the early and late time energies may indicate the presence of two outflows launched at separate times. A scenario where the accretion rate and subsequent jet launch is significantly delayed relative to peak optical light may explain the presence of a more energetic outflow launched several hundred days post-discovery \citep[e.g.,][]{Metzger_2022}. It is also possible that accretion starts earlier and the jet is launched only once the accretion rate falls below some Eddington threshold \citep{Giannios_2011,Pasham_2018}.

If the radio emission in ASASSN-19bt is in fact due to a relativistic jet launched away from our line of sight, then one might expect the temporal evolution to follow that of AT2018hyz. 
In Figures \ref{fig:Ek_gamma} and \ref{fig:density_profile}, we show that ASASSN-19bt exhibits very similar radii, energy, outflow velocity, and density to AT2018hyz using the off-axis jet model. In contrast, the 5 GHz light curve for ASASSN-19bt appears remarkably different from that of AT2018hyz (see Figure \ref{fig:lit_lc_radio}). Furthermore the radio $F_p$ for ASASSN-19bt is plateauing, which is counter-intuitive for an off-axis jet that has not yet decelerated to non-relativistic speeds. Given the freely expanding jet and steep density profile, this evolution may be still acceptable, but is not indefinitely sustainable. An important prediction of the off-axis model (if true) is that $\Gamma$ should decline in future epochs and the radio luminosity should increase again as the jet decelerates to $\Gamma\sim1$.


\subsection{X-ray Emission}\label{sec:xray_disc}

As seen in Figure \ref{fig:lit_lc}, TDEs exhibit a diverse range of X-ray luminosities. ASASSN-19bt has the faintest known X-ray emission of any optically-selected TDE at $\delta t\lesssim50$ days and remains faint in the X-rays to late times ($\delta t\sim700$ days). Initially-faint X-ray emission followed by a brightening phase has often been explained as evidence for delayed formation of the accretion disk \citep[e.g.,][]{Gezari_2017, liu_22}. However, \citet{guolo_23} showed that X-ray obscuration may be a more likely cause for this behavior. Many models invoke some form of X-ray suppression via the reprocessing of X-ray emission into lower-energy emission by optically thick gas surrounding the black hole. In particular, some have proposed models where the degree of X-ray reprocessing is orientation dependent \citep[see][]{Dai_2018, Thomsen_2022, Parkinson_2022}. In the context of these models, the reprocessing layer is produced by super-Eddington disk outflows, which would result in early time X-ray suppression corresponding to when the accretion rate was highest. For TDEs with low inclination angles, where the accretion disk is viewed face-on, the expected reprocessing is minimal. However, for high inclination angles, where the disk is viewed edge-on, the reprocessing of X-ray emission is expected to dominate, leading to fainter emission. Under this scenario, one would expect the most X-ray dim TDEs to be those with the highest inclination angles.

The presence of the cooling break identified in \S \ref{sec:cool} suggests that the synchrotron emission seen in the radio under-predicts ASASSN-19bt's X-ray emission if extrapolated to higher frequencies, requiring an additional emission component. In Swift J1644+57, one model for the X-rays at early times is inverse Compton emission from the jet, which produces a rather hard spectrum \citep{Crumley_2016}. For ASASSN-19bt, the measured photon index $\Gamma$ is significantly harder than the X-rays of most optically-selected TDEs \citep{guolo_23}. Therefore, one possibility is that the X-rays we see in ASASSN-19bt could be inverse Compton emission.

Alternately, the observed X-rays could be intrinsically softer X-ray emission that is highly absorbed. In this case, ASASSN-19bt's exceptionally faint X-ray emission might imply that we are viewing this system nearly perfectly edge-on, which would also be consistent with the most plausible relativistic jet model for the radio emission, where the jet is launched maximally off-axis ($\sim 90^\circ$) with respect to our line of sight. However, our X-ray spectral modeling found no evidence for intrinsic X-ray absorption, although the limits are not tight. The apparent lack of X-ray absorption by neutral or partially ionized material may be caused by the X-ray obscuring material in the edge-on disk already being fully ionized. 

\section{Conclusions}\label{sec:conc}
We present the results of our radio and X-ray monitoring of the TDE ASASSN-19bt, spanning nearly four years after the onset of the optical flare. Radio emission was first detected shortly after the optical discovery and continued to rise for years afterward, with the peak radio flux density plateauing $\delta t\gtrsim1$ years post-discovery. In contrast, ASASSN-19bt displays very little activity in the X-rays; its early X-ray emission was among the least luminous of any optically-selected TDE and no X-rays have been detected since approximately 225 days post-discovery.

The location of the cooling break suggests that the X-ray emission is not synchrotron in origin. We find that ASASSN-19bt's harder X-ray spectrum hints at inverse Compton emission from a possible jet or heavily absorbed softer X-ray emission being viewed through a reprocessing layer (we see no evidence for absorption in the X-ray spectrum, but the constraints are modest due to the faintness of the source). In the latter case, the lack of luminous X-ray emission may be a result of viewing the accretion disk edge-on, as orientation-dependent X-ray reprocessing models align with ASASSN-19bt's dim X-ray emission.

We employed two models to describe the possible origins of the radio emission, a non-relativistic spherical blast wave and a relativistic jet launched off-axis from our line of sight. We find that a spherical outflow model would imply a continuous energy rise in the outflow from $\sim10^{46}$ erg to $\sim10^{49}$ erg with speeds consistent with $v\approx 0.016c-0.09c$. A bulk outflow velocity of $\beta\approx0.05$ aligns with expected velocities for accretion-driven winds, unbound debris, or collision-induced outflows. Notably, all of these non-relativistic scenarios fall short in explaining the observed outflow acceleration within the first 186 days. In contrast, the off-axis relativistic model for a jet with $\theta_j = 1/\Gamma$ instead implies that the outflow expands at speeds corresponding to $\Gamma\approx10$ and takes on two periods of roughly constant energy at $\sim 2 \times 10^{50}$ erg to $\sim 7 \times 10^{51}$ erg for the most off-axis cases ($\theta_\text{obs} \gtrsim 60^\circ$). If we impose a fixed jet opening angle of $\theta_j = 10^\circ$, we instead see decreasing energy at early times and a roughly constant energy of $E\approx 10^{52}$ erg at late times in the maximally off-axis case. The peak flux density and self-absorption frequency both increase $\sim1$ year post-optical peak, resulting in discontinuities in the magnetic field and density profiles surrounding the SMBH for both models.

We find that the off-axis relativistic jet model alleviates some of the concerns presented by the simple Newtonian model (e.g.~the large increase in the energy), but also has some aspects that appear physically implausible (e.g.~the overall energy budget). Neither model appears to provide a complete explanation for the radio emission and evolution, suggesting that the true emission geometry is likely more complex than either of the two limiting cases considered here, possibly involving multiple outflow mechanisms operating on different timescales. A key takeaway from this analysis is that more holistic TDE models are needed, incorporating information from emission across the electromagnetic spectrum. 

ASASSN-19bt now becomes part of the collection of TDEs demonstrating unusual radio emission in the later stages. Undertaking extended radio observations of these TDEs, particularly at late times ($\sim$years post-disruption), will play a pivotal role in understanding the mechanisms behind the unusual late time radio emission displayed by a growing number of TDEs. We particularly emphasize the importance of Very Long Baseline Interferometry (VLBI) observations for nearby TDEs, as they are crucial for directly testing the off-axis relativistic jet hypothesis. The perpendicular motion of the jet relative to our line of sight makes VLBI observations of centroid motion a definitive smoking gun for relativistic jets within these systems \citep{ARP_2018}. Careful interpretation of the physical parameters derived from the equipartition process may be needed on an event-by-event basis to properly distinguish between truly non-relativistic outflows and highly off-axis jets, with implications for the overall occurrence rate of powerful jets in TDEs.

\section*{Acknowledgements}

The Australia Telescope Compact Array is part of the Australia Telescope National Facility (\url{https://ror.org/05qajvd42}) which is funded by the Australian Government for operation as a National Facility managed by CSIRO. We acknowledge the Gomeroi people as the Traditional Owners of the Observatory site. The National Radio Astronomy Observatory (NRAO) is a facility of the National Science Foundation operated under cooperative agreement by Associated Universities, Inc. C.~T.~C. and K.~D.~A. acknowledge support provided by the NSF through award SOSPA9-007 from the NRAO and award AST-2307668. This work was partially supported by the Australian government through the Australian Research Council’s Discovery Projects funding scheme (DP200102471).  The Berger Time-Domain Group at Harvard is supported by NSF and NASA grants.
F.D.C. acknowledges support from the DGAPA/PAPIIT grant 113424.

This paper makes use of the following ALMA data: ADS/JAO.ALMA\#2018.1.01766.T. ALMA is a partnership of ESO (representing its member states), NSF (USA) and NINS (Japan), together with NRC (Canada), NSTC and ASIAA (Taiwan), and KASI (Republic of Korea), in cooperation with the Republic of Chile. The Joint ALMA Observatory is operated by ESO, AUI/NRAO and NAOJ. We acknowledge the use of public data from the Swift data archive. This research is partially based on observations made by the Chandra X-ray Observatory under program DDT-20708675 (obsIDs 22182 and 22183; PI
Alexander) and software provided by the Chandra X-ray Center (CXC) in the application package CIAO.

\clearpage

\appendix

\section{Total Flux SED Fits and Equipartition Analysis}
\label{appendix:sed}
\begin{figure*}[!ht]
    \centering
    \includegraphics[width = \textwidth]{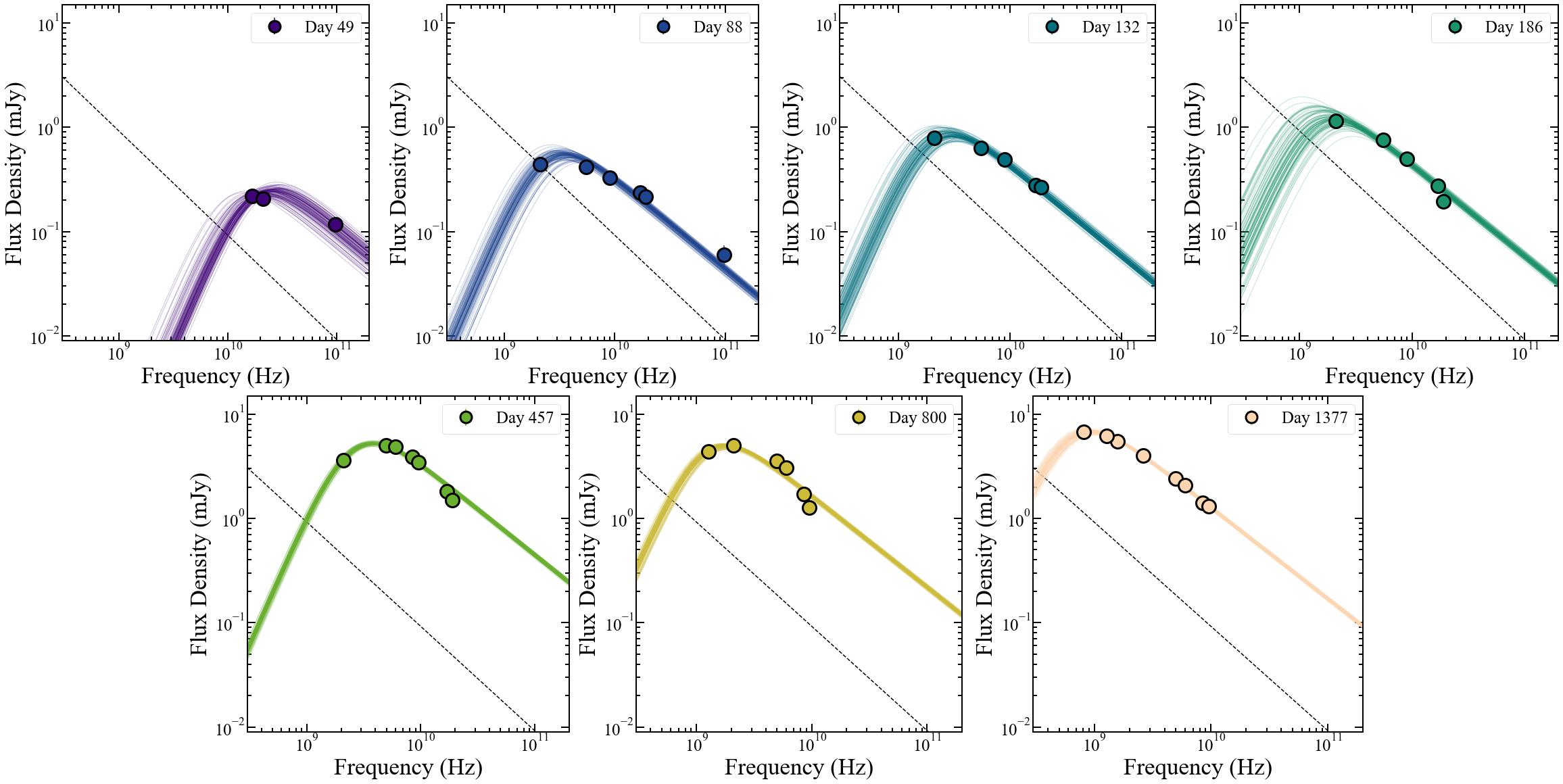}
    \caption{Radio spectral energy distribution fits to the observed ATCA, ALMA, and MeerKAT data assuming no quiescent radio component. The dashed line indicates the assumed quiescent component that we subtracted during our main analysis. The solid lines indicate a representative sample of SED fits from the MCMC modeling.}
    \label{fig:1_orig}
\end{figure*}

\begin{figure*}[!ht]
    \centering
    \includegraphics[width = \textwidth]{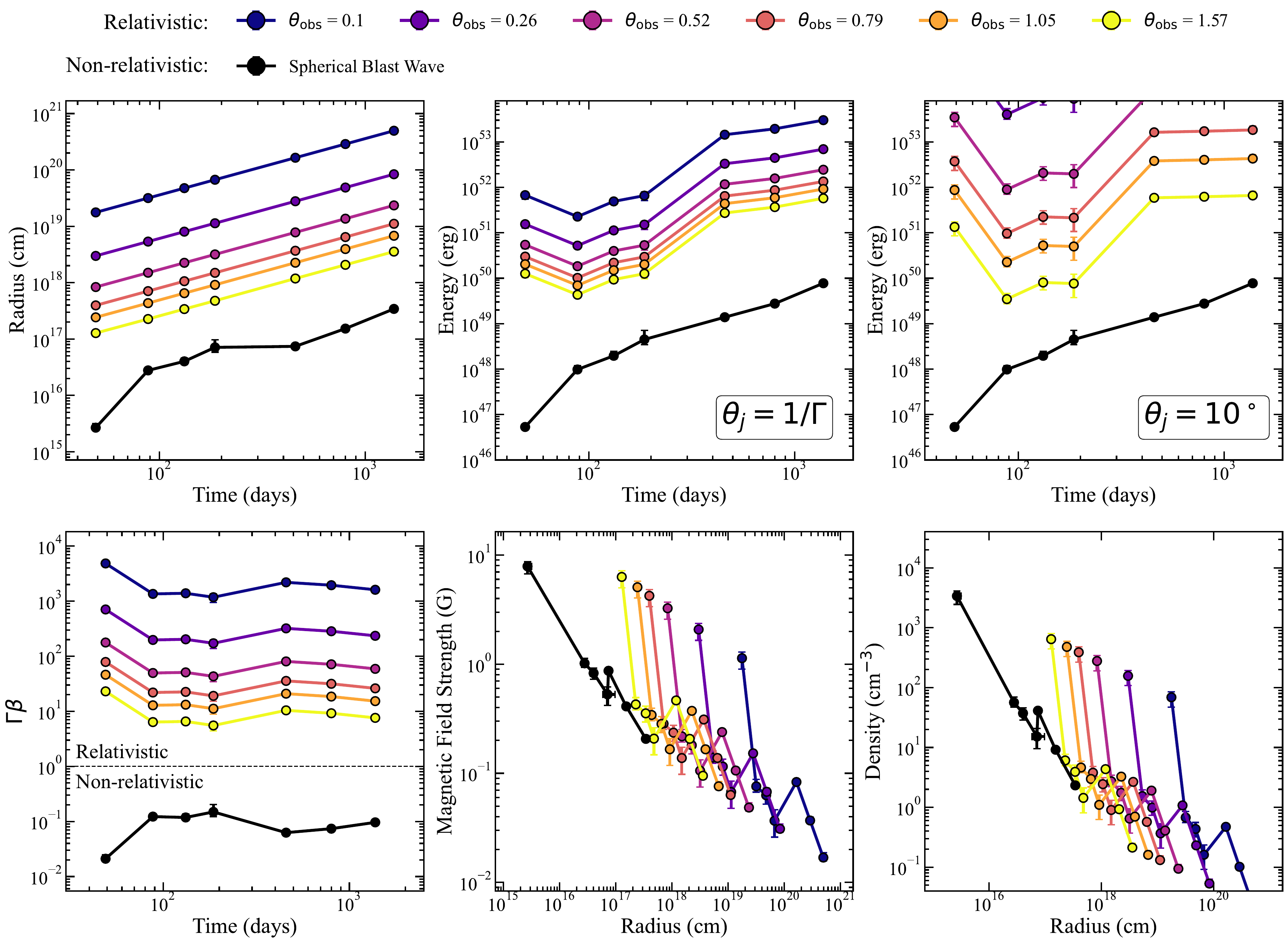}
    \caption{The temporal and radial dependencies of the physical parameters derived from our  equipartition calculations assuming no quiescent radio emission. In each panel we show the results for the non-relativistic outflow model presented in \citet{Barniol_Duran_2013} (black circles) and the relativistic solution proposed in \citet{matsumoto} (indigo - yellow circles) for a set of off-axis viewing angles. We show the radius of the emitting region as a function of time (\textit{Top, Left}), the outflow kinetic energy as a function of time for a jet with $\theta_j=1/\Gamma$ (\textit{Top, Middle}),  the outflow kinetic energy as a function of time assuming a fixed jet opening angle $\theta_j=10^\circ$ (\textit{Top, Right}), the velocity evolution of outflow (\textit{Bottom, Left}), the radial profile of the magnetic field (\textit{Bottom, Middle}), and the radial profile of the number density of electrons in the emitting region (\textit{Bottom, Right}). The error bars on the data correspond to 1 standard deviation computed using a Markov Chain Monte Carlo approach.}
    \label{fig:eq_jet_orig}
\end{figure*}

As described in Section \ref{sec:archival_radio}, we detect archival radio emission from the host galaxy of ASASSN-19bt that predates the TDE. We therefore opted to subtract off a quiescent emission component scaled to this detection and assumed to be constant in time, modeling only the transient emission component. For completeness, here we show the observed SEDs without subtracting any host component, their model fits, and the extracted outflow parameters using the same methods outlined in Section \S $\ref{sec:analysis}$ (Figures \ref{fig:1_orig},  \ref{fig:eq_jet_orig}, and Table \ref{tab:eb}). We note that these results do not alter the basic conclusions of our analysis. In particular, at late times the assumed host galaxy emission component is much fainter than the transient emission component at all frequencies; thus, the parameters at late times are virtually unchanged.

\begin{figure*}
    \centering
    \includegraphics[width = \textwidth]{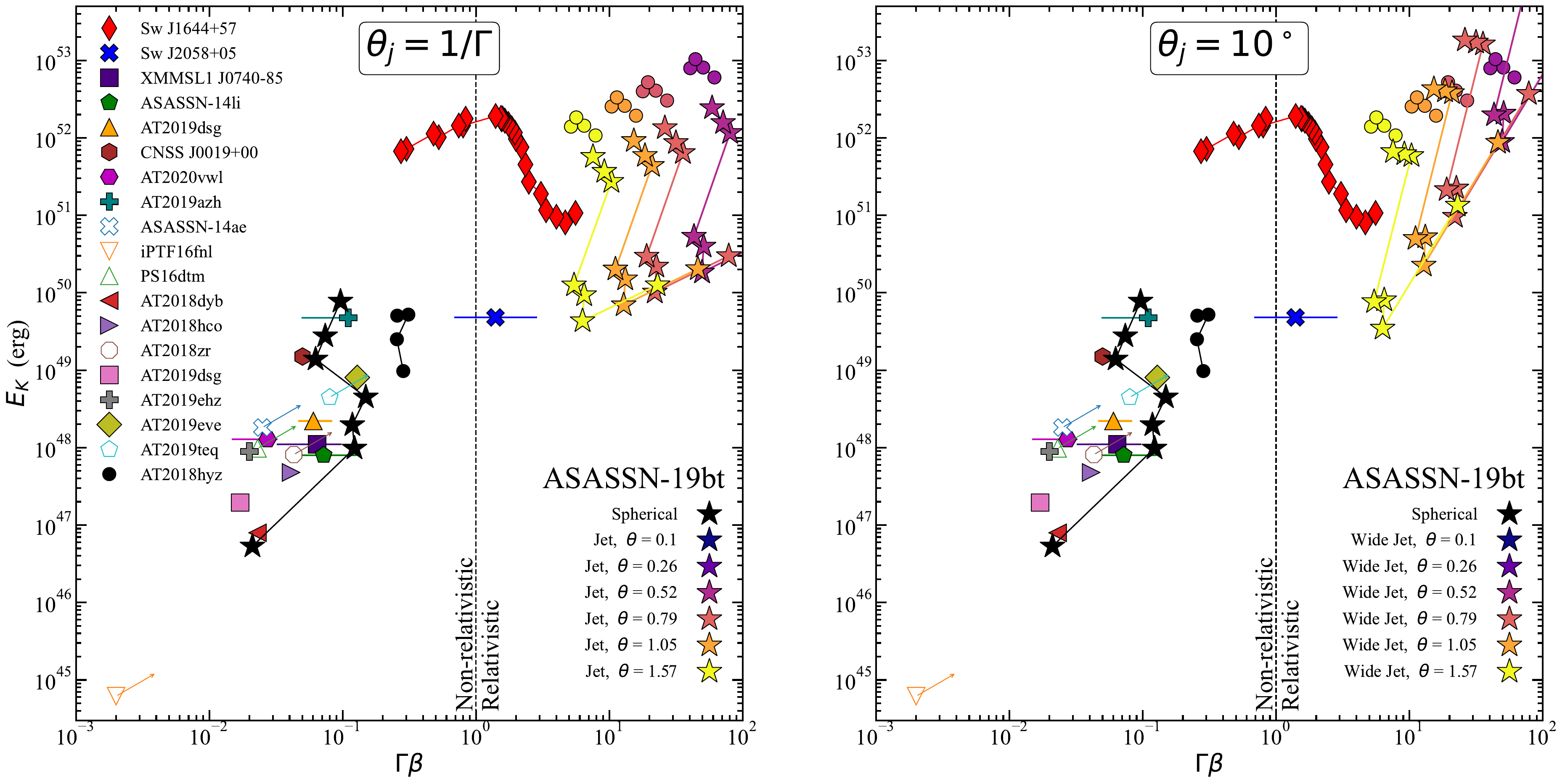}
    \caption{The outflow kinetic energies and velocities inferred from radio emission of known TDEs in context with the inferred energy evolution of ASASSN-19bt from the total flux SED fits. The non-relativistic spherical blast wave model is represented in black, while off-axis relativistic jet model is color-coded from indigo to yellow depending on the viewing angle (ASASSN-19bt is denoted by stars, and AT2018 is denoted by circles). We show the relativistic model for a jet opening angle of $\theta_j=1/\Gamma$ (\textit{Left}) and a fixed opening angle of $\theta_j=10^\circ$ (\textit{Right}). The data shown for AT2018hyz are from \citet{AT2018hyz} and \citet{matsumoto}. The remaining data are from Sw J1644+57, \citet{Zauderer_2011}, \citet{Eftekhari}, \citet{Cendes_2021}; AT2019dsg, \citealt{at2019dsg,yvette_2023}; ASASSN-14li, \citet{alexander_2016}; AT2020vwl, \citet{AT2020vwl}; AT2019azh, \citet{AT2019azh}; CNSS J0019+00, \citet{CNSS}; Sw J2058+05 , \citet{cenko_2012}; all other data are from \citet{yvette_2023}}
    \label{fig:Ek_gamma_orig}
\end{figure*}

\begin{table}[!h]
\centering
    \begin{tabular}{cccc}
    \hline\hline
        $\theta_{\rm rad}$ & $\theta_{\rm deg}$ & $\epsilon_{B,\rm transient}$ & $\epsilon_{B,\rm total flux}$ \\
        \hline

        \multicolumn{4}{c}{\textit{Off-Axis Relativistic Jet}}   \\
        0.1     &     6$^\circ$    &   0.0003  & 0.0016    \\
        0.26    &     15$^\circ$   &   0.001   & 0.004 \\
        0.52    &     30$^\circ$   &   0.002   & 0.008 \\
        0.79    &     45$^\circ$   &   0.003   & 0.012 \\
        1.05    &     60$^\circ$   &   0.003   & 0.016 \\
        1.57    &     90$^\circ$   &   0.005   & 0.022 \\
        \hline
        \multicolumn{4}{c}{\textit{Non-relativistic Spherical Outflow}}   \\
        - & - & 0.2 & 0.65 \\
    \hline \hline
    \end{tabular}
    \caption{Inferred values for $\epsilon_B$ assuming different outflow geometries.    \label{tab:eb}}
\end{table}

\bibliographystyle{aasjournal}
\bibliography{mybib} 


\label{lastpage}
\end{document}

%% file: authors.tex
\author[0000-0003-0528-202X]{Collin T. Christy}
\altaffiliation{E-mail: collinchristy@arizona.edu}
\affiliation{Department of Astronomy/Steward Observatory, 933 North Cherry Avenue, Rm. N204, 
Tucson, AZ 85721-0065, USA}

\author[0000-0002-8297-2473]{Kate D. Alexander}
\affiliation{Department of Astronomy/Steward Observatory, 933 North Cherry Avenue, Rm. N204, 
Tucson, AZ 85721-0065, USA}

\author[0000-0001-7007-6295]{Yvette Cendes}
\affiliation{Center for Astrophysics $\vert$ Harvard \& Smithsonian, Cambridge, MA 02138, USA}

\author[0000-0002-7706-5668]{Ryan Chornock}
\affiliation{Department of Astronomy, University of California, Berkeley, CA 94720-3411, USA}

\author[0000-0003-1792-2338]{Tanmoy Laskar}
\affiliation{Department of Physics \& Astronomy, University of Utah, Salt Lake City, UT 84112, USA}
\affiliation{Department of Astrophysics/IMAPP, Radboud University, PO Box 9010, 6500 GL Nijmegen, The Netherlands}

\author[0000-0003-4768-7586]{Raffaella Margutti}
\affil{Department of Astronomy, University of California, Berkeley, CA 94720-3411, USA}
\affil{Department of Physics, University of California, Berkeley, CA 94720-7300, USA}

\author[0000-0002-9392-9681]{Edo Berger}
\affiliation{Center for Astrophysics $\vert$ Harvard \& Smithsonian, Cambridge, MA 02138, USA}

\author[0000-0002-0592-4152]{Michael Bietenholz}
\affiliation{Department of Physics and Astronomy, York University, 4700 Keele St., Toronto, M3J 1P3, Ontario, Canada}

\author[0000-0001-5126-6237]{Deanne Coppejans}
\affiliation{Department of Physics, University of Warwick, Gibbet Hill Road, Coventry CV4 7AL, UK}

\author[0000-0002-3137-4633]{Fabio De Colle}
\affiliation{Instituto de Ciencias Nucleares, Universidad Nacional Aut\'{o}noma de M\'{e}xico, Apartado Postal 70-264, 04510 M\'{e}xico, CDMX, Mexico}

\author[0000-0003-0307-9984]{Tarraneh Eftekhari}
\altaffiliation{NHFP Einstein Fellow}
\affiliation{Center for Interdisciplinary Exploration and Research in Astrophysics (CIERA) and Department of Physics and Astronomy, Northwestern University, Evanston, IL
60208, USA}

\author[0000-0001-9206-3460]{Thomas W.-S. Holoien}
\affiliation{The Observatories of the Carnegie Institution for Science, 813 Santa Barbara St., Pasadena, CA 91101, USA}

\author[0000-0002-9350-6793]{Tatsuya Matsumoto}
\affiliation{Department of Astronomy, Kyoto University, Kitashirakawa-Oiwake-cho, Sakyo-ku, Kyoto, 606-8502, Japan}
\affiliation{Hakubi Center, Kyoto University, Yoshida-honmachi, Sakyo-ku, Kyoto, 606-8501, Japan}

\author[0000-0003-3124-2814]{James C. A. Miller-Jones}
\affiliation{International Centre for Radio Astronomy Research, Curtin University, GPO Box U1987, Perth, WA 6845, Australia}

\author[0000-0003-2558-3102]{Enrico Ramirez-Ruiz}
\affiliation{Department of Astronomy and Astrophysics, UCO/Lick Observatory, University of California, 1156 High Street, Santa Cruz, CA 95064, USA}

\author[0000-0002-4912-2477]{Richard Saxton}
\affiliation{Operations Department, Telespazio UK for ESA, European Space Astronomy Centre, 28691 Villanueva de la Ca\~{n}ada,  Spain}

\author[0000-0002-3859-8074]{Sjoert van Velzen}
\affiliation{Leiden Observatory, Leiden University, PO Box 9513, NL-2300 RA Leiden, the Netherlands}

\author[0000-0002-7721-8660]{Mark Wieringa}
\affiliation{CSIRO Space and Astronomy, PO Box 76, Epping NSW 1710, Australia}

%% file: main.bbl
\begin{thebibliography}{}
\expandafter\ifx\csname natexlab\endcsname\relax\def\natexlab#1{#1}\fi
\providecommand{\url}[1]{\href{#1}{#1}}
\providecommand{\dodoi}[1]{doi:~\href{http://doi.org/#1}{\nolinkurl{#1}}}
\providecommand{\doeprint}[1]{\href{http://ascl.net/#1}{\nolinkurl{http://ascl.net/#1}}}
\providecommand{\doarXiv}[1]{\href{https://arxiv.org/abs/#1}{\nolinkurl{https://arxiv.org/abs/#1}}}

\bibitem[{{Alexander} {et~al.}(2016){Alexander}, {Berger}, {Guillochon},
  {Zauderer}, \& {Williams}}]{alexander_2016}
{Alexander}, K.~D., {Berger}, E., {Guillochon}, J., {Zauderer}, B.~A., \&
  {Williams}, P.~K.~G. 2016, \apjl, 819, L25,
  \dodoi{10.3847/2041-8205/819/2/L25}

\bibitem[{{Alexander} {et~al.}(2020){Alexander}, {van Velzen}, {Horesh}, \&
  {Zauderer}}]{alexander2020}
{Alexander}, K.~D., {van Velzen}, S., {Horesh}, A., \& {Zauderer}, B.~A. 2020,
  \ssr, 216, 81, \dodoi{10.1007/s11214-020-00702-w}

\bibitem[{{Anderson} {et~al.}(2020){Anderson}, {Mooley}, {Hallinan}, {Dong},
  {Phinney}, {Horesh}, {Bourke}, {Cenko}, {Frail}, {Kulkarni}, \&
  {Myers}}]{CNSS}
{Anderson}, M.~M., {Mooley}, K.~P., {Hallinan}, G., {et~al.} 2020, \apj, 903,
  116, \dodoi{10.3847/1538-4357/abb94b}

\bibitem[{{Andreoni} {et~al.}(2022){Andreoni}, {Coughlin}, {Perley}, {Yao},
  {Lu}, {Cenko}, {Kumar}, {Anand}, {Ho}, {Kasliwal}, {de Ugarte Postigo},
  {Sagu{\'e}s-Carracedo}, {Schulze}, {Kann}, {Kulkarni}, {Sollerman}, {Tanvir},
  {Rest}, {Izzo}, {Somalwar}, {Kaplan}, {Ahumada}, {Anupama}, {Auchettl},
  {Barway}, {Bellm}, {Bhalerao}, {Bloom}, {Bremer}, {Bulla}, {Burns},
  {Campana}, {Chandra}, {Charalampopoulos}, {Cooke}, {D'Elia}, {Das}, {Dobie},
  {Ag{\"u}{\'\i} Fern{\'a}ndez}, {Freeburn}, {Fremling}, {Gezari}, {Goode},
  {Graham}, {Hammerstein}, {Karambelkar}, {Kilpatrick}, {Kool}, {Krips},
  {Laher}, {Leloudas}, {Levan}, {Lundquist}, {Mahabal}, {Medford}, {Miller},
  {M{\"o}ller}, {Mooley}, {Nayana}, {Nir}, {Pang}, {Paraskeva}, {Perley},
  {Petitpas}, {Pursiainen}, {Ravi}, {Ridden-Harper}, {Riddle}, {Rigault},
  {Rodriguez}, {Rusholme}, {Sharma}, {Smith}, {Stein}, {Th{\"o}ne},
  {Tohuvavohu}, {Valdes}, {van Roestel}, {Vergani}, {Wang}, \&
  {Zhang}}]{Andreoni_2022}
{Andreoni}, I., {Coughlin}, M.~W., {Perley}, D.~A., {et~al.} 2022, \nat, 612,
  430, \dodoi{10.1038/s41586-022-05465-8}

\bibitem[{{Barniol Duran} {et~al.}(2013){Barniol Duran}, Nakar, \&
  Piran}]{Barniol_Duran_2013}
{Barniol Duran}, R., Nakar, E., \& Piran, T. 2013, The Astrophysical Journal,
  772, 78, \dodoi{10.1088/0004-637X/772/1/78}

\bibitem[{{Beniamini} {et~al.}(2023){Beniamini}, {Piran}, \&
  {Matsumoto}}]{Beniamini_2023}
{Beniamini}, P., {Piran}, T., \& {Matsumoto}, T. 2023, \mnras, 524, 1386,
  \dodoi{10.1093/mnras/stad1950}

\bibitem[{{Berger} {et~al.}(2012){Berger}, {Zauderer}, {Pooley}, {Soderberg},
  {Sari}, {Brunthaler}, \& {Bietenholz}}]{berger_2012}
{Berger}, E., {Zauderer}, A., {Pooley}, G.~G., {et~al.} 2012, \apj, 748, 36,
  \dodoi{10.1088/0004-637X/748/1/36}

\bibitem[{{Bondi}(1952)}]{Bondi_1952}
{Bondi}, H. 1952, \mnras, 112, 195, \dodoi{10.1093/mnras/112.2.195}

\bibitem[{{Bonnerot} \& {Lu}(2020)}]{bonnerot_2020}
{Bonnerot}, C., \& {Lu}, W. 2020, \mnras, 495, 1374,
  \dodoi{10.1093/mnras/staa1246}

\bibitem[{{Brown} {et~al.}(2017){Brown}, {Levan}, {Stanway}, {Kr{\"u}hler},
  {Tanvir}, {Davies}, {Fruchter}, {Cenko}, \& {Metzger}}]{Brown_2017}
{Brown}, G.~C., {Levan}, A.~J., {Stanway}, E.~R., {et~al.} 2017, \mnras, 472,
  4469, \dodoi{10.1093/mnras/stx2193}

\bibitem[{{Burrows} {et~al.}(2005){Burrows}, {Hill}, {Nousek}, {Kennea},
  {Wells}, {Osborne}, {Abbey}, {Beardmore}, {Mukerjee}, {Short}, {Chincarini},
  {Campana}, {Citterio}, {Moretti}, {Pagani}, {Tagliaferri}, {Giommi},
  {Capalbi}, {Tamburelli}, {Angelini}, {Cusumano}, {Br{\"a}uninger}, {Burkert},
  \& {Hartner}}]{Burrows05}
{Burrows}, D.~N., {Hill}, J.~E., {Nousek}, J.~A., {et~al.} 2005, \ssr, 120,
  165, \dodoi{10.1007/s11214-005-5097-2}

\bibitem[{{CASA Team} {et~al.}(2022){CASA Team}, {Bean}, {Bhatnagar}, {Castro},
  {Donovan Meyer}, {Emonts}, {Garcia}, {Garwood}, {Golap}, {Gonzalez Villalba},
  {Harris}, {Hayashi}, {Hoskins}, {Hsieh}, {Jagannathan}, {Kawasaki},
  {Keimpema}, {Kettenis}, {Lopez}, {Marvil}, {Masters}, {McNichols},
  {Mehringer}, {Miel}, {Moellenbrock}, {Montesino}, {Nakazato}, {Ott}, {Petry},
  {Pokorny}, {Raba}, {Rau}, {Schiebel}, {Schweighart}, {Sekhar}, {Shimada},
  {Small}, {Steeb}, {Sugimoto}, {Suoranta}, {Tsutsumi}, {van Bemmel},
  {Verkouter}, {Wells}, {Xiong}, {Szomoru}, {Griffith}, {Glendenning}, \&
  {Kern}}]{CASA_team}
{CASA Team}, {Bean}, B., {Bhatnagar}, S., {et~al.} 2022, \pasp, 134, 114501,
  \dodoi{10.1088/1538-3873/ac9642}

\bibitem[{{Cendes} {et~al.}(2021{\natexlab{a}}){Cendes}, {Alexander}, {Berger},
  {Eftekhari}, {Williams}, \& {Chornock}}]{at2019dsg}
{Cendes}, Y., {Alexander}, K.~D., {Berger}, E., {et~al.} 2021{\natexlab{a}},
  \apj, 919, 127, \dodoi{10.3847/1538-4357/ac110a}

\bibitem[{{Cendes} {et~al.}(2021{\natexlab{b}}){Cendes}, {Eftekhari}, {Berger},
  \& {Polisensky}}]{Cendes_2021}
{Cendes}, Y., {Eftekhari}, T., {Berger}, E., \& {Polisensky}, E.
  2021{\natexlab{b}}, \apj, 908, 125, \dodoi{10.3847/1538-4357/abd323}

\bibitem[{{Cendes} {et~al.}(2022){Cendes}, {Berger}, {Alexander}, {Gomez},
  {Hajela}, {Chornock}, {Laskar}, {Margutti}, {Metzger}, {Bietenholz},
  {Brethauer}, \& {Wieringa}}]{AT2018hyz}
{Cendes}, Y., {Berger}, E., {Alexander}, K.~D., {et~al.} 2022, \apj, 938, 28,
  \dodoi{10.3847/1538-4357/ac88d0}

\bibitem[{{Cendes} {et~al.}(2023){Cendes}, {Berger}, {Alexander}, {Chornock},
  {Margutti}, {Metzger}, {Wieringa}, {Bietenholz}, {Hajela}, {Laskar}, {Stroh},
  \& {Terreran}}]{yvette_2023}
---. 2023, arXiv e-prints, arXiv:2308.13595, \dodoi{10.48550/arXiv.2308.13595}

\bibitem[{{Cenko} {et~al.}(2012){Cenko}, {Krimm}, {Horesh}, {Rau}, {Frail},
  {Kennea}, {Levan}, {Holland}, {Butler}, {Quimby}, {Bloom}, {Filippenko},
  {Gal-Yam}, {Greiner}, {Kulkarni}, {Ofek}, {Olivares E.}, {Schady},
  {Silverman}, {Tanvir}, \& {Xu}}]{cenko_2012}
{Cenko}, S.~B., {Krimm}, H.~A., {Horesh}, A., {et~al.} 2012, \apj, 753, 77,
  \dodoi{10.1088/0004-637X/753/1/77}

\bibitem[{{Chevalier}(1998)}]{1998chevalier}
{Chevalier}, R.~A. 1998, \apj, 499, 810, \dodoi{10.1086/305676}

\bibitem[{{Condon} {et~al.}(2002){Condon}, {Cotton}, \&
  {Broderick}}]{Condon_2002}
{Condon}, J.~J., {Cotton}, W.~D., \& {Broderick}, J.~J. 2002, \aj, 124, 675,
  \dodoi{10.1086/341650}

\bibitem[{{Crumley} {et~al.}(2016){Crumley}, {Lu}, {Santana}, {Hern{\'a}ndez},
  {Kumar}, \& {Markoff}}]{Crumley_2016}
{Crumley}, P., {Lu}, W., {Santana}, R., {et~al.} 2016, \mnras, 460, 396,
  \dodoi{10.1093/mnras/stw967}

\bibitem[{{Dai} {et~al.}(2018){Dai}, {McKinney}, {Roth}, {Ramirez-Ruiz}, \&
  {Miller}}]{Dai_2018}
{Dai}, L., {McKinney}, J.~C., {Roth}, N., {Ramirez-Ruiz}, E., \& {Miller},
  M.~C. 2018, \apjl, 859, L20, \dodoi{10.3847/2041-8213/aab429}

\bibitem[{{Duffell} \& {Laskar}(2018)}]{duffell_2018}
{Duffell}, P.~C., \& {Laskar}, T. 2018, \apj, 865, 94,
  \dodoi{10.3847/1538-4357/aadb9c}

\bibitem[{{Eftekhari} {et~al.}(2018){Eftekhari}, {Berger}, {Zauderer},
  {Margutti}, \& {Alexander}}]{Eftekhari}
{Eftekhari}, T., {Berger}, E., {Zauderer}, B.~A., {Margutti}, R., \&
  {Alexander}, K.~D. 2018, \apj, 854, 86, \dodoi{10.3847/1538-4357/aaa8e0}

\bibitem[{Eftekhari {et~al.}(2024)Eftekhari, Tchekhovskoy, Alexander, Berger,
  Chornock, Laskar, Margutti, Yao, Cendes, Gomez, Hajela, \&
  Pasham}]{eftekhari_2024}
Eftekhari, T., Tchekhovskoy, A., Alexander, K.~D., {et~al.} 2024, Late-time
  X-ray Observations of the Jetted Tidal Disruption Event AT2022cmc: The
  Relativistic Jet Shuts Off.
\newblock \doarXiv{2404.10036}

\bibitem[{{Evans} {et~al.}(2009){Evans}, {Beardmore}, {Page}, {Osborne},
  {O'Brien}, {Willingale}, {Starling}, {Burrows}, {Godet}, {Vetere}, {Racusin},
  {Goad}, {Wiersema}, {Angelini}, {Capalbi}, {Chincarini}, {Gehrels}, {Kennea},
  {Margutti}, {Morris}, {Mountford}, {Pagani}, {Perri}, {Romano}, \&
  {Tanvir}}]{Evans09}
{Evans}, P.~A., {Beardmore}, A.~P., {Page}, K.~L., {et~al.} 2009, \mnras, 397,
  1177, \dodoi{10.1111/j.1365-2966.2009.14913.x}

\bibitem[{{Foreman-Mackey} {et~al.}(2013){Foreman-Mackey}, {Hogg}, {Lang}, \&
  {Goodman}}]{emcee}
{Foreman-Mackey}, D., {Hogg}, D.~W., {Lang}, D., \& {Goodman}, J. 2013, \pasp,
  125, 306, \dodoi{10.1086/670067}

\bibitem[{{Gehrels} {et~al.}(2004){Gehrels}, {Chincarini}, {Giommi}, {Mason},
  {Nousek}, {Wells}, {White}, {Barthelmy}, {Burrows}, {Cominsky}, {Hurley},
  {Marshall}, {M{\'e}sz{\'a}ros}, {Roming}, {Angelini}, {Barbier}, {Belloni},
  {Campana}, {Caraveo}, {Chester}, {Citterio}, {Cline}, {Cropper}, {Cummings},
  {Dean}, {Feigelson}, {Fenimore}, {Frail}, {Fruchter}, {Garmire}, {Gendreau},
  {Ghisellini}, {Greiner}, {Hill}, {Hunsberger}, {Krimm}, {Kulkarni}, {Kumar},
  {Lebrun}, {Lloyd-Ronning}, {Markwardt}, {Mattson}, {Mushotzky}, {Norris},
  {Osborne}, {Paczynski}, {Palmer}, {Park}, {Parsons}, {Paul}, {Rees},
  {Reynolds}, {Rhoads}, {Sasseen}, {Schaefer}, {Short}, {Smale}, {Smith},
  {Stella}, {Tagliaferri}, {Takahashi}, {Tashiro}, {Townsley}, {Tueller},
  {Turner}, {Vietri}, {Voges}, {Ward}, {Willingale}, {Zerbi}, \&
  {Zhang}}]{Gehrels04}
{Gehrels}, N., {Chincarini}, G., {Giommi}, P., {et~al.} 2004, \apj, 611, 1005,
  \dodoi{10.1086/422091}

\bibitem[{{Gezari} {et~al.}(2017){Gezari}, {Cenko}, \& {Arcavi}}]{Gezari_2017}
{Gezari}, S., {Cenko}, S.~B., \& {Arcavi}, I. 2017, \apjl, 851, L47,
  \dodoi{10.3847/2041-8213/aaa0c2}

\bibitem[{{Giannios} \& {Metzger}(2011)}]{Giannios_2011}
{Giannios}, D., \& {Metzger}, B.~D. 2011, \mnras, 416, 2102,
  \dodoi{10.1111/j.1365-2966.2011.19188.x}

\bibitem[{{Goldstein} {et~al.}(2016){Goldstein}, {Connaughton}, {Briggs}, \&
  {Burns}}]{Goldstein_2016}
{Goldstein}, A., {Connaughton}, V., {Briggs}, M.~S., \& {Burns}, E. 2016, \apj,
  818, 18, \dodoi{10.3847/0004-637X/818/1/18}

\bibitem[{{Goodwin} {et~al.}(2022){Goodwin}, {van Velzen}, {Miller-Jones},
  {Mummery}, {Bietenholz}, {Wederfoort}, {Hammerstein}, {Bonnerot}, {Hoffmann},
  \& {Yan}}]{AT2019azh}
{Goodwin}, A.~J., {van Velzen}, S., {Miller-Jones}, J.~C.~A., {et~al.} 2022,
  \mnras, 511, 5328, \dodoi{10.1093/mnras/stac333}

\bibitem[{{Goodwin} {et~al.}(2023{\natexlab{a}}){Goodwin}, {Alexander},
  {Miller-Jones}, {Bietenholz}, {van Velzen}, {Anderson}, {Berger}, {Cendes},
  {Chornock}, {Coppejans}, {Eftekhari}, {Gezari}, {Laskar}, {Ramirez-Ruiz}, \&
  {Saxton}}]{AT2020vwl}
{Goodwin}, A.~J., {Alexander}, K.~D., {Miller-Jones}, J.~C.~A., {et~al.}
  2023{\natexlab{a}}, \mnras, 522, 5084, \dodoi{10.1093/mnras/stad1258}

\bibitem[{{Goodwin} {et~al.}(2023{\natexlab{b}}){Goodwin}, {Miller-Jones}, {van
  Velzen}, {Bietenholz}, {Greenland}, {Cenko}, {Gezari}, {Horesh}, {Sivakoff},
  {Yan}, {Yu}, \& {Zhang}}]{at2020opy}
{Goodwin}, A.~J., {Miller-Jones}, J.~C.~A., {van Velzen}, S., {et~al.}
  2023{\natexlab{b}}, \mnras, 518, 847, \dodoi{10.1093/mnras/stac3127}

\bibitem[{{Govreen-Segal} {et~al.}(2021){Govreen-Segal}, {Nakar}, \&
  {Levinson}}]{Govreen-Segal_2021}
{Govreen-Segal}, T., {Nakar}, E., \& {Levinson}, A. 2021, \apj, 907, 113,
  \dodoi{10.3847/1538-4357/abd180}

\bibitem[{Granot \& Sari(2002)}]{Granot_2002}
Granot, J., \& Sari, R. 2002, The Astrophysical Journal, 568, 820,
  \dodoi{10.1086/338966}

\bibitem[{{Guolo} {et~al.}(2023){Guolo}, {Gezari}, {Yao}, {van Velzen},
  {Hammerstein}, {Cenko}, \& {Tokayer}}]{guolo_23}
{Guolo}, M., {Gezari}, S., {Yao}, Y., {et~al.} 2023, arXiv e-prints,
  arXiv:2308.13019, \dodoi{10.48550/arXiv.2308.13019}

\bibitem[{{Hills}(1975)}]{hills_1975}
{Hills}, J.~G. 1975, \nat, 254, 295, \dodoi{10.1038/254295a0}

\bibitem[{{Holoien} {et~al.}(2019){Holoien}, {Vallely}, {Auchettl}, {Stanek},
  {Kochanek}, {French}, {Prieto}, {Shappee}, {Brown}, {Fausnaugh}, {Dong},
  {Thompson}, {Bose}, {Neustadt}, {Cacella}, {Brimacombe}, {Kendurkar},
  {Beaton}, {Boutsia}, {Chomiuk}, {Connor}, {Morrell}, {Newman}, {Rudie},
  {Shishkovksy}, \& {Strader}}]{holoien_2019}
{Holoien}, T. W.~S., {Vallely}, P.~J., {Auchettl}, K., {et~al.} 2019, \apj,
  883, 111, \dodoi{10.3847/1538-4357/ab3c66}

\bibitem[{{Horesh} {et~al.}(2021{\natexlab{a}}){Horesh}, {Cenko}, \&
  {Arcavi}}]{horesh_15oi}
{Horesh}, A., {Cenko}, S.~B., \& {Arcavi}, I. 2021{\natexlab{a}}, Nature
  Astronomy, 5, 491, \dodoi{10.1038/s41550-021-01300-8}

\bibitem[{{Horesh} {et~al.}(2021{\natexlab{b}}){Horesh}, {Sfaradi}, {Fender},
  {Green}, {Williams}, \& {Bright}}]{iPTF16fnl}
{Horesh}, A., {Sfaradi}, I., {Fender}, R., {et~al.} 2021{\natexlab{b}}, \apjl,
  920, L5, \dodoi{10.3847/2041-8213/ac25fe}

\bibitem[{{Kalberla} {et~al.}(2005){Kalberla}, {Burton}, {Hartmann}, {Arnal},
  {Bajaja}, {Morras}, \& {P{\"o}ppel}}]{Kalberla05}
{Kalberla}, P.~M.~W., {Burton}, W.~B., {Hartmann}, D., {et~al.} 2005, \aap,
  440, 775, \dodoi{10.1051/0004-6361:20041864}

\bibitem[{{Kochanek} {et~al.}(2017){Kochanek}, {Shappee}, {Stanek}, {Holoien},
  {Thompson}, {Prieto}, {Dong}, {Shields}, {Will}, {Britt}, {Perzanowski}, \&
  {Pojma{\'n}ski}}]{Kochanek_2017}
{Kochanek}, C.~S., {Shappee}, B.~J., {Stanek}, K.~Z., {et~al.} 2017, \pasp,
  129, 104502, \dodoi{10.1088/1538-3873/aa80d9}

\bibitem[{{Krolik} {et~al.}(2016){Krolik}, {Piran}, {Svirski}, \&
  {Cheng}}]{krolik_2016}
{Krolik}, J., {Piran}, T., {Svirski}, G., \& {Cheng}, R.~M. 2016, \apj, 827,
  127, \dodoi{10.3847/0004-637X/827/2/127}

\bibitem[{{Lei} {et~al.}(2016){Lei}, {Yuan}, {Zhang}, \& {Wang}}]{IGR_Lei}
{Lei}, W.-H., {Yuan}, Q., {Zhang}, B., \& {Wang}, D. 2016, \apj, 816, 20,
  \dodoi{10.3847/0004-637X/816/1/20}

\bibitem[{{Li} {et~al.}(2018){Li}, {Li}, {Huang}, {Geng}, {Yu}, \&
  {Song}}]{li_2018}
{Li}, B., {Li}, L.-B., {Huang}, Y.-F., {et~al.} 2018, \apjl, 859, L3,
  \dodoi{10.3847/2041-8213/aac2c5}

\bibitem[{{Liu} {et~al.}(2022){Liu}, {Dou}, {Chen}, \& {Shen}}]{liu_22}
{Liu}, X.-L., {Dou}, L.-M., {Chen}, J.-H., \& {Shen}, R.-F. 2022, \apj, 925,
  67, \dodoi{10.3847/1538-4357/ac33a9}

\bibitem[{{Mangano} {et~al.}(2016){Mangano}, {Burrows}, {Sbarufatti}, \&
  {Cannizzo}}]{Mangano_2016}
{Mangano}, V., {Burrows}, D.~N., {Sbarufatti}, B., \& {Cannizzo}, J.~K. 2016,
  \apj, 817, 103, \dodoi{10.3847/0004-637X/817/2/103}

\bibitem[{{Margalit} \& {Quataert}(2021)}]{Margalit_2021}
{Margalit}, B., \& {Quataert}, E. 2021, \apjl, 923, L14,
  \dodoi{10.3847/2041-8213/ac3d97}

\bibitem[{{Margutti} {et~al.}(2013){Margutti}, {Zaninoni}, {Bernardini},
  {Chincarini}, {Pasotti}, {Guidorzi}, {Angelini}, {Burrows}, {Capalbi},
  {Evans}, {Gehrels}, {Kennea}, {Mangano}, {Moretti}, {Nousek}, {Osborne},
  {Page}, {Perri}, {Racusin}, {Romano}, {Sbarufatti}, {Stafford}, \&
  {Stamatikos}}]{Margutti13}
{Margutti}, R., {Zaninoni}, E., {Bernardini}, M.~G., {et~al.} 2013, \mnras,
  428, 729, \dodoi{10.1093/mnras/sts066}

\bibitem[{{Matsumoto} \& {Metzger}(2023)}]{AT2022cmc_matsumoto_metzger2023}
{Matsumoto}, T., \& {Metzger}, B.~D. 2023, \mnras, 522, 4028,
  \dodoi{10.1093/mnras/stad1182}

\bibitem[{{Matsumoto} \& {Piran}(2023)}]{matsumoto}
{Matsumoto}, T., \& {Piran}, T. 2023, \mnras, 522, 4565,
  \dodoi{10.1093/mnras/stad1269}

\bibitem[{{Matsumoto} {et~al.}(2022){Matsumoto}, {Piran}, \&
  {Krolik}}]{matsumoto_19dsg}
{Matsumoto}, T., {Piran}, T., \& {Krolik}, J.~H. 2022, \mnras, 511, 5085,
  \dodoi{10.1093/mnras/stac382}

\bibitem[{{Mattila} {et~al.}(2018){Mattila}, {P{\'e}rez-Torres}, {Efstathiou},
  {Mimica}, {Fraser}, {Kankare}, {Alberdi}, {Aloy}, {Heikkil{\"a}}, {Jonker},
  {Lundqvist}, {Mart{\'\i}-Vidal}, {Meikle}, {Romero-Ca{\~n}izales}, {Smartt},
  {Tsygankov}, {Varenius}, {Alonso-Herrero}, {Bondi}, {Fransson},
  {Herrero-Illana}, {Kangas}, {Kotak}, {Ram{\'\i}rez-Olivencia},
  {V{\"a}is{\"a}nen}, {Beswick}, {Clements}, {Greimel}, {Harmanen},
  {Kotilainen}, {Nandra}, {Reynolds}, {Ryder}, {Walton}, {Wiik}, \&
  {{\"O}stlin}}]{ARP_2018}
{Mattila}, S., {P{\'e}rez-Torres}, M., {Efstathiou}, A., {et~al.} 2018,
  Science, 361, 482, \dodoi{10.1126/science.aao4669}

\bibitem[{{Metzger}(2022)}]{Metzger_2022}
{Metzger}, B.~D. 2022, \apjl, 937, L12, \dodoi{10.3847/2041-8213/ac90ba}

\bibitem[{{Metzger} {et~al.}(2012){Metzger}, {Giannios}, \&
  {Mimica}}]{Metzger_2012}
{Metzger}, B.~D., {Giannios}, D., \& {Mimica}, P. 2012, \mnras, 420, 3528,
  \dodoi{10.1111/j.1365-2966.2011.20273.x}

\bibitem[{{Panaitescu} \& {Kumar}(2002)}]{Panaitescu_2002}
{Panaitescu}, A., \& {Kumar}, P. 2002, \apj, 571, 779, \dodoi{10.1086/340094}

\bibitem[{{Parkinson} {et~al.}(2022){Parkinson}, {Knigge}, {Matthews}, {Long},
  {Higginbottom}, {Sim}, \& {Mangham}}]{Parkinson_2022}
{Parkinson}, E.~J., {Knigge}, C., {Matthews}, J.~H., {et~al.} 2022, \mnras,
  510, 5426, \dodoi{10.1093/mnras/stac027}

\bibitem[{{Pasham} \& {van Velzen}(2018)}]{Pasham_2018}
{Pasham}, D.~R., \& {van Velzen}, S. 2018, \apj, 856, 1,
  \dodoi{10.3847/1538-4357/aab361}

\bibitem[{{Pasham} {et~al.}(2023){Pasham}, {Lucchini}, {Laskar}, {Gompertz},
  {Srivastav}, {Nicholl}, {Smartt}, {Miller-Jones}, {Alexander}, {Fender},
  {Smith}, {Fulton}, {Dewangan}, {Gendreau}, {Coughlin}, {Rhodes}, {Horesh},
  {van Velzen}, {Sfaradi}, {Guolo}, {Castro Segura}, {Aamer}, {Anderson},
  {Arcavi}, {Brennan}, {Chambers}, {Charalampopoulos}, {Chen}, {Clocchiatti},
  {de Boer}, {Dennefeld}, {Ferrara}, {Galbany}, {Gao}, {Gillanders}, {Goodwin},
  {Gromadzki}, {Huber}, {Jonker}, {Joshi}, {Kara}, {Killestein}, {Kosec},
  {Kocevski}, {Leloudas}, {Lin}, {Margutti}, {Mattila}, {Moore},
  {M{\"u}ller-Bravo}, {Ngeow}, {Oates}, {Onori}, {Pan}, {Perez-Torres}, {Rani},
  {Remillard}, {Ridley}, {Schulze}, {Sheng}, {Shingles}, {Smith}, {Steiner},
  {Wainscoat}, {Wevers}, \& {Yang}}]{AT2022cmc_pasham_2023}
{Pasham}, D.~R., {Lucchini}, M., {Laskar}, T., {et~al.} 2023, Nature Astronomy,
  7, 88, \dodoi{10.1038/s41550-022-01820-x}

\bibitem[{{Perlman} {et~al.}(2017){Perlman}, {Meyer}, {Wang}, {Yuan},
  {Henriksen}, {Irwin}, {Krause}, {Wiegert}, {Murphy}, {Heald}, \&
  {Dettmar}}]{perlman_2017}
{Perlman}, E.~S., {Meyer}, E.~T., {Wang}, Q.~D., {et~al.} 2017, \apj, 842, 126,
  \dodoi{10.3847/1538-4357/aa71b1}

\bibitem[{{Perlman} {et~al.}(2022){Perlman}, {Meyer}, {Wang}, {Yuan},
  {Henriksen}, {Irwin}, {Li}, {Wiegert}, {Li}, \& {Yang}}]{perlman_2022}
---. 2022, \apj, 925, 143, \dodoi{10.3847/1538-4357/ac3bba}

\bibitem[{{Pushkarev} {et~al.}(2017){Pushkarev}, {Kovalev}, {Lister}, \&
  {Savolainen}}]{Pushkarev_2017}
{Pushkarev}, A.~B., {Kovalev}, Y.~Y., {Lister}, M.~L., \& {Savolainen}, T.
  2017, \mnras, 468, 4992, \dodoi{10.1093/mnras/stx854}

\bibitem[{{Rees}(1988)}]{Rees1998}
{Rees}, M.~J. 1988, \nat, 333, 523, \dodoi{10.1038/333523a0}

\bibitem[{{Riccio} {et~al.}(2023){Riccio}, {Yang}, {Ma{\l}ek}, {Boquien},
  {Junais}, {Pistis}, {Hamed}, {Grespan}, {Paolillo}, \&
  {Torbaniuk}}]{Riccio_2023}
{Riccio}, G., {Yang}, G., {Ma{\l}ek}, K., {et~al.} 2023, \aap, 678, A164,
  \dodoi{10.1051/0004-6361/202346857}

\bibitem[{{Ricker} {et~al.}(2015){Ricker}, {Winn}, {Vanderspek}, {Latham},
  {Bakos}, {Bean}, {Berta-Thompson}, {Brown}, {Buchhave}, {Butler}, {Butler},
  {Chaplin}, {Charbonneau}, {Christensen-Dalsgaard}, {Clampin}, {Deming},
  {Doty}, {De Lee}, {Dressing}, {Dunham}, {Endl}, {Fressin}, {Ge}, {Henning},
  {Holman}, {Howard}, {Ida}, {Jenkins}, {Jernigan}, {Johnson}, {Kaltenegger},
  {Kawai}, {Kjeldsen}, {Laughlin}, {Levine}, {Lin}, {Lissauer}, {MacQueen},
  {Marcy}, {McCullough}, {Morton}, {Narita}, {Paegert}, {Palle}, {Pepe},
  {Pepper}, {Quirrenbach}, {Rinehart}, {Sasselov}, {Sato}, {Seager},
  {Sozzetti}, {Stassun}, {Sullivan}, {Szentgyorgyi}, {Torres}, {Udry}, \&
  {Villasenor}}]{Ricker_2015}
{Ricker}, G.~R., {Winn}, J.~N., {Vanderspek}, R., {et~al.} 2015, Journal of
  Astronomical Telescopes, Instruments, and Systems, 1, 014003,
  \dodoi{10.1117/1.JATIS.1.1.014003}

\bibitem[{{Rouco Escorial} {et~al.}(2023){Rouco Escorial}, {Fong}, {Berger},
  {Laskar}, {Margutti}, {Schroeder}, {Rastinejad}, {Cornish}, {Popp}, {Lally},
  {Nugent}, {Paterson}, {Metzger}, {Chornock}, {Alexander}, {Cendes}, \&
  {Eftekhari}}]{Rouco_2023}
{Rouco Escorial}, A., {Fong}, W., {Berger}, E., {et~al.} 2023, \apj, 959, 13,
  \dodoi{10.3847/1538-4357/acf830}

\bibitem[{{Sari} {et~al.}(1998){Sari}, {Piran}, \& {Narayan}}]{Sari_1998}
{Sari}, R., {Piran}, T., \& {Narayan}, R. 1998, \apjl, 497, L17,
  \dodoi{10.1086/311269}

\bibitem[{{Sault} {et~al.}(1995){Sault}, {Teuben}, \& {Wright}}]{Miriad}
{Sault}, R.~J., {Teuben}, P.~J., \& {Wright}, M.~C.~H. 1995, in Astronomical
  Society of the Pacific Conference Series, Vol.~77, Astronomical Data Analysis
  Software and Systems IV, ed. R.~A. {Shaw}, H.~E. {Payne}, \& J.~J.~E.
  {Hayes}, 433, \dodoi{10.48550/arXiv.astro-ph/0612759}

\bibitem[{{Sfaradi} {et~al.}(2022){Sfaradi}, {Horesh}, {Fender}, {Green},
  {Williams}, {Bright}, \& {Schulze}}]{sfaradi_19azh}
{Sfaradi}, I., {Horesh}, A., {Fender}, R., {et~al.} 2022, \apj, 933, 176,
  \dodoi{10.3847/1538-4357/ac74bc}

\bibitem[{{Sfaradi} {et~al.}(2024){Sfaradi}, {Beniamini}, {Horesh}, {Piran},
  {Bright}, {Rhodes}, {Williams}, {Fender}, {Leung}, {Murphy}, \&
  {Green}}]{Sfaradi_2024}
{Sfaradi}, I., {Beniamini}, P., {Horesh}, A., {et~al.} 2024, \mnras, 527, 7672,
  \dodoi{10.1093/mnras/stad3717}

\bibitem[{{Shappee} {et~al.}(2014){Shappee}, {Prieto}, {Grupe}, {Kochanek},
  {Stanek}, {De Rosa}, {Mathur}, {Zu}, {Peterson}, {Pogge}, {Komossa}, {Im},
  {Jencson}, {Holoien}, {Basu}, {Beacom}, {Szczygie{\l}}, {Brimacombe},
  {Adams}, {Campillay}, {Choi}, {Contreras}, {Dietrich}, {Dubberley},
  {Elphick}, {Foale}, {Giustini}, {Gonzalez}, {Hawkins}, {Howell}, {Hsiao},
  {Koss}, {Leighly}, {Morrell}, {Mudd}, {Mullins}, {Nugent}, {Parrent},
  {Phillips}, {Pojmanski}, {Rosing}, {Ross}, {Sand}, {Terndrup}, {Valenti},
  {Walker}, \& {Yoon}}]{Shappee}
{Shappee}, B.~J., {Prieto}, J.~L., {Grupe}, D., {et~al.} 2014, \apj, 788, 48,
  \dodoi{10.1088/0004-637X/788/1/48}

\bibitem[{{Strubbe} \& {Quataert}(2009)}]{Strubbe}
{Strubbe}, L.~E., \& {Quataert}, E. 2009, \mnras, 400, 2070,
  \dodoi{10.1111/j.1365-2966.2009.15599.x}

\bibitem[{{Tchekhovskoy} {et~al.}(2014){Tchekhovskoy}, {Metzger}, {Giannios},
  \& {Kelley}}]{Tchekhovskoy}
{Tchekhovskoy}, A., {Metzger}, B.~D., {Giannios}, D., \& {Kelley}, L.~Z. 2014,
  \mnras, 437, 2744, \dodoi{10.1093/mnras/stt2085}

\bibitem[{{Thomsen} {et~al.}(2022){Thomsen}, {Kwan}, {Dai}, {Wu}, {Roth}, \&
  {Ramirez-Ruiz}}]{Thomsen_2022}
{Thomsen}, L.~L., {Kwan}, T.~M., {Dai}, L., {et~al.} 2022, \apjl, 937, L28,
  \dodoi{10.3847/2041-8213/ac911f}

\bibitem[{{van Velzen} {et~al.}(2013){van Velzen}, {Frail}, {K{\"o}rding}, \&
  {Falcke}}]{van_Velzen_2023}
{van Velzen}, S., {Frail}, D.~A., {K{\"o}rding}, E., \& {Falcke}, H. 2013,
  \aap, 552, A5, \dodoi{10.1051/0004-6361/201220426}

\bibitem[{{Williams} {et~al.}(2019){Williams}, {Allers}, {Biller}, \&
  {Vos}}]{peeling}
{Williams}, P.~K.~G., {Allers}, K.~N., {Biller}, B.~A., \& {Vos}, J. 2019,
  Research Notes of the American Astronomical Society, 3, 110,
  \dodoi{10.3847/2515-5172/ab35d5}

\bibitem[{{Yao} {et~al.}(2024){Yao}, {Lu}, {Harrison}, {Kulkarni}, {Gezari},
  {Guolo}, {Cenko}, \& {Ho}}]{Yao_2024}
{Yao}, Y., {Lu}, W., {Harrison}, F., {et~al.} 2024, \apj, 965, 39,
  \dodoi{10.3847/1538-4357/ad2b6b}

\bibitem[{{Zauderer} {et~al.}(2011){Zauderer}, {Berger}, {Soderberg}, {Loeb},
  {Narayan}, {Frail}, {Petitpas}, {Brunthaler}, {Chornock}, {Carpenter},
  {Pooley}, {Mooley}, {Kulkarni}, {Margutti}, {Fox}, {Nakar}, {Patel},
  {Volgenau}, {Culverhouse}, {Bietenholz}, {Rupen}, {Max-Moerbeck}, {Readhead},
  {Richards}, {Shepherd}, {Storm}, \& {Hull}}]{Zauderer_2011}
{Zauderer}, B.~A., {Berger}, E., {Soderberg}, A.~M., {et~al.} 2011, \nat, 476,
  425, \dodoi{10.1038/nature10366}

\end{thebibliography}
